\documentclass[acmsmall,screen,authorversion,nonacm]{acmart}

\usepackage{multirow}

\AtBeginDocument{%
  }

\copyrightyear{2026}
\acmYear{2026}
\setcopyright{cc}
\setcctype{by}
\acmConference[CHI '26]{Proceedings of the 2026 CHI Conference on Human Factors in Computing Systems}{April 13--17, 2026}{Barcelona, Spain}
\acmBooktitle{Proceedings of the 2026 CHI Conference on Human Factors in Computing Systems (CHI '26), April 13--17, 2026, Barcelona, Spain}
\acmDOI{10.1145/3772318.3793150}
\acmISBN{979-8-4007-2278-3/2026/04}

\begin{document}

\title{Chasing Meaning and/or Insight? A Survey on Evaluation Practices at the Intersection of Visualization and the Humanities}

\author{Alejandro Benito-Santos}
\affiliation{%
  \department{Department for Computer Languages and Systems}
  \institution{National Distance Education University (UNED)}
  \city{Madrid}
  \country{Spain}
}
\email{al.benito@lsi.uned.es}
\orcid{0000-0001-5317-6390}

\author{Florian Windhager}
\affiliation{%
  \department{Dept. for Arts and Cultural Studies}
  \institution{University for Continuing Education}
  \city{Krems}
  \country{Austria}
}
\email{florian.windhager@donau-uni.ac.at}
\orcid{0000-0002-5170-2243}

\author{Aida Horaniet Ibañez}
\affiliation{%
  \institution{University of Luxembourg}
  \city{Esch-sur-Alzette}
  \country{Luxembourg}
}
\email{aida.horanietibanez@uni.lu}
\orcid{0000-0003-1346-8494}

\author{Rabea Kleymann}
\affiliation{%
  \institution{University of Technology Chemnitz}
  \city{Chemnitz}
  \country{Germany}
}
\email{rabea.kleymann@phil.tu-chemnitz.de}
\orcid{0000-0003-3856-2685}

\author{Alfie Abdul-Rahman}
\affiliation{%
  \department{Department of Informatics}
  \institution{King's College London}
  \city{London}
  \country{United Kingdom}
}
\email{alfie.abdulrahman@kcl.ac.uk}
\orcid{0000-0002-6257-876X}

\author{Eva Mayr}
\affiliation{%
  \department{Dept. for Arts and Cultural Studies}
  \institution{University for Continuing Education}
  \city{Krems}
  \country{Austria}
}
\email{eva.mayr@donau-uni.ac.at}
\orcid{0000-0001-8402-5990}

\renewcommand{\shortauthors}{Benito-Santos et al.}
\renewcommand{\shorttitle}{Chasing Meaning and/or Insight?}


\begin{abstract}
The intersection of visualization and the humanities (VIS*H) is marked by a tension between chasing analytical ``insight'' and interpretive ``meaning.'' The effectiveness of visualization techniques hinges on established evaluation frameworks that assess both analytical utility and communicative efficacy, creating a potential mismatch with the non-positivist, interpretive aims of humanities scholarship. To examine how this tension manifests in practice, we systematically surveyed 171 VIS*H design studies to analyze their evaluation workflows and rigor according to standard practice. Our findings reveal recurring flaws, such as an over-reliance on monomethod approaches, and show that higher-quality evaluations emerge from workflows that effectively triangulate diverse evidence. From these findings, we derive recommendations to refine quality and validation criteria for humanities visualizations, and juxtapose them to ongoing critical debates in the field, ultimately arguing for a paradigm shift that can reconcile the advantages of established validation techniques with the interpretive depth required for humanistic inquiry.
\end{abstract}





\maketitle

\section{Introduction}
\label{introduction}

Digital methods have entered the working repertoire of the arts and humanities, and visualization has become one of their most prominent expressions. 
\footnote{In the absence of an unbiased shorthand for this field (cf. \cite{bradley_visualization_2018}), we will refer to it as VIS*H---an acronym that accommodates multiple operationalizations and constellations within a tripolar community of practice, spanning visualization research (VIS), digital humanities (DH), and traditional humanities (H). These include prominent unilateral delivery scenarios such as VIS$\rightarrow$DH and critical responses VIS$\leftarrow$H, but also more balanced constellations characterized by mutual learning cycles (VIS$\rightleftarrows$DH$\rightleftarrows$H) which represent the field's inherent promise.} This development represents a significant shift within the humanities, as it complements traditional language- and text-based reasoning methods with more formalized and analytical approaches, and also with visual and interactive means of exploring and interpreting complex datasets \cite{dork2011information, janicke_visual_2017}. Yet, at the core of this shift lies a fundamental tension: while the ``insight'' derived from pattern discovery or outlier identification is a core strength of visualization \cite{North2006, stasko2014value}, the ``meaning'' central to the humanities arises from interpretive and discursive acts that require a careful and critical engagement with cultural materials, considering not only their explicit content but also the multiple possible readings related to their motivations and historical contexts \cite{drucker_humanities_2011, lamqaddam_introducing_2020}---a process that poses a significant challenge for digital formalization \cite{drucker_visualization_2020, correll2019counting}. 

As such, evaluating these visualizations has become a critical area of research \cite{bradley_visualization_2018, Hinrichs:DSH:2018}, necessitating both a survey of existing evaluation approaches and an exploration of the unique affordances and demands of humanities data and of related scholarly communities. This paper presents an extensive study on the current state of evaluation methods in visualizing humanities data, highlighting prevalent methodologies, but also mapping out future challenges and advancements for this rapidly expanding field.

We consider the relevance of this topic to derive from the unique nature of humanities data, conceived here as digital representations that carry both observational and interpretive properties of \emph{cultural materials}. These materials---and cultures as a whole---can be conceptualized as ``human creations that define what is real and important'' for those who create, preserve, or share them \cite[p.627]{kimmel2006culture}. Whether as texts, images, ideas, music, narratives, or movies, cultural materials carry both \emph{meaning} and \emph{value}, from small and mundane to collective and historical levels. From a sociological point of view, they cause and catalyze both social cohesion but also conflict by delighting, informing, and connecting collectives---and thus also distancing or antagonizing others~\cite{jackson2024worldwide, inglehart2005christian}. As highly diverse and complex subject matters---whether historical or contemporary---they document the diversity and development of human thinking, feeling, and striving---motivating their study and preservation due to their ``aesthetic, historic, scientific, or social value''~\cite{icomos1982florence-charta}. 

Consequently, datasets about such materials and topics often encompass complex conceptual, qualitative, and value-laden dimensions, in addition to temporal, spatial, or relational characteristics, which are relatively easy to process by formal analytical means. We consider this specific mixture of interpretive and formally tractable features---and related methodological tensions---to be a core reason why the VIS*H field appears as a notably challenging and promising `application field', requiring innovative, context-specific, and (self-)critical approaches to visualization design. Applying standard datafication, formalization, and visualization methods to this field can significantly enhance the work of humanities experts (a stance we will refer to as the ``discovery paradigm''), but is also frequently seen as not enough (a perspective we label the ``discursive paradigm''):\footnote{We adopt these labels as provisional shorthand, with ``discursive'' referring to humanities discourse as the site where the core methodological movements of \emph{interpretation} and \emph{critique} intertwine (see Sec.\ref{sec:challenges}).}

i) \emph{The Discovery Paradigm}: The integration of visualization methods in humanities research is often lauded for facilitating a formally deeper and quantitatively more comprehensive understanding of cultural and historical datasets, democratizing access to complex information---from individual artifacts to large corpora---and for making them more comprehensible and engaging to a broader audience. The effectiveness, comprehensibility, and success of these visualizations, however, are contingent upon rigorous evaluation methods that can assess both their analytical utility (including factors like ease of use and task effectiveness) and their ability to communicate insights to diverse audiences.

ii) \emph{The Discursive Paradigm}: Visualizations have, however, also attracted a significant amount of  `meta-evaluative' commentary and critique in the humanities, scrutinizing their epistemological and methodological value at various levels. From the outset, debates have emerged around the question of whether visualizations can productively contribute to humanities scholarship in their current form $(\mathrm{VIS}\rightarrow H)$---including adaptations to the field through user-centered design $(\mathrm{VIS}\rightleftarrows H)$---or whether they must undergo significant transformation, to address the substantial challenges\footnote{One of these challenges is the need to reconcile computational and visualization-based approaches with the dominant interpretive and discursive practices in the humanities. Other substantial challenges, as elaborated in \autoref{sec:challenges}, involve representing the provenance and uncertainty inherent in cultural data and aligning visual designs with the critical and theoretical frameworks that guide humanistic inquiry.} arising from its focus on \textit{cultural, semiotic, value-laden objects, topics, and materials} $(\mathrm{VIS}\leftarrow H)$ \cite{bradley_visualization_2018, drucker_humanities_2011}. As such, questions persist as to whether visualization and other digital methods should be considered a) as legitimate, core research methods, b) as ancillary, supporting instruments working orthogonally to established humanities methods, or c) as distractions, deformations, or `Trojan horses' that should instead be avoided \cite{correll2019counting, drucker_humanities_2011, lamqaddam_introducing_2020, bentkowska2013bought}.

In turn, this paper examines both the landscape of the growing exploratory VIS*H field (VIS$\rightarrow$H) and its validation methods (often aligned with the discovery paradigm) and synthesizes critical and transformative arguments (i.e., VIS$\leftarrow$H, aligned with the discursive paradigm), to discuss their implications for future evaluation endeavors. In doing so, it aims to contribute to the development of more effective, accessible, and insightful visualization tools in humanities research, while also preparing the visualization field for future revisions of some of its quality and validation criteria, ultimately improving our understanding and meaning-making of complex cultural phenomena (cf. \cite{meyer2019criteria, berret_iceberg_2025, lamqaddam_introducing_2020}). 

The impulse to address these questions in a more focused and systematic fashion emerged from a Dagstuhl Seminar on ``Visualization and the Humanities: Towards a Shared Research Agenda'' \cite{drucker_et_al:DagRep.13.9.137}, where scholars and practitioners from visualization and the humanities gathered to discuss the state of this working area and related future challenges. One of the working groups soon zoomed in on two main questions, which guided our work: a) What is the state of the art in evaluating work in VIS*H, and b) How far do evaluation efforts engage with the specific challenges of intersectional applications in data visualization?

In the following, we summarize related work (\autoref{sec:relatedwork}) and the methodology guiding our survey (\autoref{methodology}). We present the categories with which we analyzed the VIS*H evaluation design space (\autoref{sec:categorization}) to assess main results of our inquiry (sections \ref{sec:summary_results} and \ref{sec:analysis_of_workflows}), and discuss specific findings (\autoref{sec:discussion}). To complement and enrich this collection of challenges for future projects, we summarize main arguments from `meta-evaluative' VIS*H debates, and discuss implications for extended validation approaches (\autoref{sec:challenges}).

\section{Related Work}
\label{sec:relatedwork}

The outlined endeavor draws from prior work around the intersection of \emph{visualization of humanities data} (as an activity distributed across digital humanities, HCI, and visualization communities) and \emph{evaluation research}.

\subsection{Demarcation of ``Humanities'' Work} 
Defining and delineating humanities scholarship is challenging due to its diverse and complex nature and due to the contested, fuzzy boundaries of this knowledge domain ~\cite{bod_new_2016,daston_history_2015,benito-santos_data-driven_2020}. At their core, the humanities study human culture, history, language, and artistic expression. 
This kind of work happens across a multitude of disciplines, whose reliable enumeration, demarcation, and overall clustering is challenging in itself, as it largely ties back to taxonomic and bureaucratic conventions handled differently in various cultural, institutional, and historical contexts \cite{during2020what_were, bod_new_2016}.\footnote{Most frequently, the ``humanities'' (whether as umbrella term or reference to a whole faculty) integrate \emph{literature}, \emph{linguistics} and \emph{language studies}, the \emph{arts} and their history (incl. visual arts (painting and sculpture) and performing arts (music, dance, theater)), \emph{philosophy}, \emph{religion}, and \emph{history} as the study of human culture, creations, and activities through time. According to historical developments, more recent fields such as cultural, ethnic, gender, or media studies are often discrete, disciplinary units of humanities faculties, as there are frequent arguments to include those parts of psychology, pedagogy, linguistics, architecture, anthropology, or sociology, working with humanities methods.} 

This definitional ambiguity is amplified within the context of the ``Digital Humanities'' (DH), which are often characterized as a `big tent' that gathers a vast array of disciplinary contributions~\cite{terras_defining_2013, bianco2012digital}. This conceptual breadth has led scholars to argue that DH resists being defined by a consistent set of theoretical concerns or research methods. Instead, it is more productively understood as a community of practice---a ``social category'' shaped by its affiliates and their scholarly interactions, rather than an ``ontological one'' with a stable disciplinary core~\cite{alvarado_digital_2012}.

\subsection{Visualizing Humanities Data}

While \emph{visualizing humanities data} represents a relatively new field of study, surveys and overview papers have already documented many of its primary areas and challenges. Various surveys and specialized studies have explored visualization research within the humanities and cultural heritage, examining visualization practices, challenges, and applications in these contexts. For example, surveys by J\"anicke et al.~\cite{janicke_close_2015, janicke_visual_2017}, Windhager et al.~\cite{Windhager:TVCG:2019}, Benito-Santos and Therón~\cite{benito-santos_data-driven_2020}, and Panagiotidou et al.~\cite{panagiotidou_communicating_2023} provide comprehensive overviews of the visualization methods applied in digital humanities and cultural heritage fields. These surveys map out major trends, but also emphasize specific requirements of humanities-centered visualization research, including the need for qualitative insights, interpretability, and transparency in visualization design.

\subsection{Evaluating Visualizations}

Evaluating visualization techniques and applications is a critical task in the field, especially for user-centered approaches aiming and claiming to ``help people carry out tasks more effectively''~\cite[p.1]{munzner2014visualization}. Notable contributions from Isenberg et al.~\cite{isenberg_systematic_2013} and Sedlmair et al.~\cite{Sedlmair:TVCG:2012} document evaluation approaches that range from controlled experiments to qualitative case studies. Researchers have identified various challenges in evaluation~\cite{Carpendale2008, Isenberg2008, Plaisant2004} and have developed distinct objectives, scenarios, and methods. More recent surveys have expanded this picture by examining how evaluation methods are distributed across different contribution types, highlighting tensions around the number and combination of evaluation techniques that are expected in a single study~\cite{Lin2024}, and by analyzing evaluation practices in domain application research, where considerations of requirements, context, and domain expertise play a pivotal role~\cite{xing_review_2025}. Most studies on visualization evaluation emphasize usability and effectiveness~\cite{dal2002evaluating, Banissi2014}, but interest in user experience is growing~\cite{saket2016beyond}. Also, insight generation is increasingly recognized as a crucial element of evaluation~\cite{North2006,saraiya2006, Smuc2009}. For example, Stasko~\cite{stasko2014value} introduced a value-driven framework to assess a system’s effectiveness in answering questions, generating insights, conveying essential data, and building user confidence.

The visualization community employs both quantitative and qualitative evaluation methods to validate these claims, such as surveys, interviews, focus groups, case studies, and controlled experiments. Carpendale~\cite{Carpendale2008} examined empirical evaluation methods, while North~\cite{North2006} proposed an insight-based approach that uses open protocols. Although visualization evaluation practices have similarities with fields such as psychology and sociology, comparative analyses between disciplines are uncommon~\cite{crisan2018how}. Despite substantial research on visual evaluation methods, cross-domain comparisons remain limited. Although some reviews focus on specific areas, such as medical visualization~\cite{Preim2018ACA}, broader comparisons among different visualization application domains are still needed. These studies emphasize the necessity of aligning evaluation methods with the goals and complexities of the visualization task, particularly in domain-specific applications.

\subsection{Evaluating Visualizations of Humanities Data}

Building on these foundations, several studies have enhanced our understanding of the challenges associated with evaluation in visualization for the (digital) humanities. Bradley et al.~\cite{bradley_visualization_2018} explore critical aspects of evaluation frameworks in VIS*H, highlighting the need for methodologies that address interpretative and contextual nuances. Benito-Santos et al.~\cite{benito-santos_evaluating_2021} and Panagiotidou et al.~\cite{panagiotidou_implicit_2021} further investigate the role of uncertainty visualization in humanities research workflows, identifying key gaps in evaluative practices. Drucker~\cite{drucker_performative_2013} focuses on best practices for assessing visualization in digital humanities contexts, presenting five core principles to guide evaluation efforts. Hinrichs et al.~\cite{Hinrichs:DSH:2018} highlight the necessity of refining evaluation criteria to better capture the effectiveness of visualization tools in supporting humanities scholarship. Finally, Lamqaddam et al.~\cite{lamqaddam_introducing_2020} on \textit{Layers of Meaning} address the challenge of closing semantic gaps between visualizations and the cultural materials they represent, proposing strategies to enhance the interpretability and communicative clarity of visual representations.

Despite these contributions---illuminating relevant aspects from different perspectives---we see a significant gap regarding surveys of research in VIS*H and related evaluation-focused studies. Although there is an expanding body of literature on visualization for the humanities and evaluation methodologies in visualization, comprehensive surveys specifically addressing evaluation within the VIS*H context have not been undertaken and published yet. Our work seeks to bridge this gap by providing a structured analysis of evaluation practices in VIS*H, synthesizing existing approaches, identifying challenges, and offering guidance for future research.

%


\section{Methodology}
\label{methodology}%

\subsection{Team Positionality and Evaluation Lens}
\label{sec:positionality}
The methodological choices documented below are shaped by the interdisciplinary composition of the author team, which spans visualization research, empirical HCI, psychology, digital humanities, cultural history, literary studies, and philosophy of knowledge. Crucially, every author possesses extensive experience serving as a reviewer for specialized journals and conferences across these domains. This mix of backgrounds informed our expectations of what constitutes methodological quality, evidential grounding, and interpretive depth in VIS*H work. Extended experience with visualization evaluation positioned some of us to attend closely to study design, sampling logic, and the consistency of methodological claims, while humanities expertise foregrounded the interpretive, contextual, and theory-dependent quality of meaning-making in the humanities.

We recognize that this constellation of perspectives sparked discussions about whether to examine VIS*H evaluation workflows primarily for interpretive richness or for systematic patterns that could be compared across studies. Reflecting on the scale of the corpus and the absence of field-wide baselines in the existing literature, we decided to identify and code quantifiable features of evaluation practice to characterize evaluation diversity at scale. This emphasis risks foregrounding forms of rigor familiar from empirical visualization research, so we combined structured quantitative coding with iterative qualitative annotation, discussion, and category refinement to prevent any single methodological stance from dominating the synthesis.

Furthermore, all authors currently work within European academic institutions, which shapes our familiarity with publications, publication norms, evaluation expectations, and VIS*H discourse. We note that evaluation practices differ across global contexts and institutional traditions, and this situatedness may have influenced how we interpreted concepts such as evaluation quality and rigor. To mitigate this, coding and synthesis were carried out collaboratively across disciplinary perspectives, and divergences in judgment were treated as prompts to refine categories rather than as errors to be resolved.

\subsection{Creation of a Seed Sample}

Given the inherently complex and multifaceted nature of the DH field, this survey employed a pragmatic approach to compile a representative sample of VIS*H literature, consciously avoiding keyword-based web searches—which are considered inadequate for capturing the comprehensive scope of the field \cite{benito-santos_data-driven_2020}. Instead, a seed corpus was developed, comprising key seminal publications within the VIS*H domain that served as the starting point to retrieve related work via literature snowballing \cite{wohlin_guidelines_2014}. Although the creation of a seed sample allowed us to scan the field's intellectual network without using keywords, we also acknowledge its primary limitation: it is not exhaustive and may overlook relevant papers that fall outside the direct citation network of our seed corpus, particularly emerging or isolated pockets of research. However, we are confident that our expert-curated seed sample provides a robust and representative snapshot of the central discourse and most influential works within the VIS*H field. The following sections describe this process in detail.

In order to build a sensible seed dataset, each of the six authors of this paper was asked to contribute five seminal publications that, in their judgment, captured the breadth of VIS*H work, resulting in a total of 30 nominees. In making their selections, the authors were also encouraged---when possible---to maintain a balanced mix of papers originating in the visualization domain and in the humanities domain. Nominees were not limited to empirical design study papers; surveys, position pieces, and other theoretically oriented contributions were equally admissible as long as they were viewed as field-defining.\footnote{Articles addressing crucial methodological or theoretical discussions in this field were included in the seed sample, but later excluded from the survey, to be analyzed for extracting future challenges only (sec.~\ref{sec:challenges}).} To mitigate potential selection bias and ensure the seed sample was not unduly influenced by the authors' own work and research networks, contributors were explicitly instructed not to nominate papers on which they were an author. Also, they were asked to favor papers with a healthy citation record whenever feasible, thereby ensuring that subsequent forward- and backward-snowballing would uncover a dense network of additional references. After collecting all the contributions and removing duplicates, the combined selection comprised 17 articles published in recognized visualization venues or journals, such as \textit{IEEE Transactions on Visualization and Computer Graphics}, \textit{Computer Graphics Forum}, or the \textit{CHI}, \textit{IEEE VIS}, and \textit{EuroVis} conferences—and in DH journals such as \textit{Digital Humanities Quarterly} or \textit{Digital Scholarship in the Humanities}. This list, presented in \autoref{tab:seed}, served as the starting point for the larger sampling pipeline described below. 

\begin{table*}[tb]
\caption{Seed publications that anchored the snowball sampling procedure.
           Fwd / Bwd are the number of works in the final sample originating from a given item in the seed, either in the forward or the backward passes, respectively.}
  \begin{tabular}{llcccc}
    \toprule
    Abbreviated Title & First author & Venue & Year & Fwd & Bwd\\
    \midrule
    Can I Believe What I See?~\cite{boyd_davis_can_2021} & Boyd Davis & ISR & 2021 & 0 & 0\\
    Visualization and the Digital Humanities~\cite{bradley_visualization_2018} & Bradley & IEEE CG\&A & 2018 & 11 & 4\\
    Feminist Data Visualization~\cite{dignazio_feminist_2016} & D’Ignazio & VIS4DH & 2016 & 9 & 0\\
    Humanities Approaches to Graphical Display~\cite{drucker_humanities_2011} & Drucker & DHQ & 2011 & 33 & 0\\
    In Defense of Sandcastles~\cite{hinrichs_defense_2019} & Hinrichs & DSH & 2019 & 2 & 0\\
    Close \& Distant Reading Survey~\cite{janicke_close_2015} & Jänicke & EuroVis STAR & 2015 & 36 & 14\\
    Interactive Visual Profiling of Musicians~\cite{janicke_interactive_2016} & Jänicke & TVCG & 2016 & 4 & 0\\
    Valuable Research for Vis \& DH~\cite{janicke_valuable_2016} & Jänicke & VIS4DH & 2016 & 8 & 3\\
    Visual Text Analysis in DH~\cite{janicke_visual_2017} & Jänicke & CGF & 2017 & 14 & 19\\
    VarifocalReader~\cite{koch_varifocalreader_2014} & Koch & TVCG & 2014 & 16 & 2\\
    Introducing Layers of Meaning~\cite{lamqaddam_introducing_2020} & Lamqaddam & TVCG & 2021 & 3 & 4\\
    Poemage~\cite{mccurdy_poemage_2016} & McCurdy & TVCG & 2016 & 15 & 2\\
    Communicating Uncertainty in DH Vis~\cite{panagiotidou_communicating_2023} & Panagiotidou & TVCG & 2023 & 1 & 9\\
    A Network Framework of Cultural History~\cite{schich_network_2014} & Schich & Science & 2014 & 3 & 0\\
    The Bohemian Bookshelf~\cite{thudt_bohemian_2012} & Thudt & CHI & 2012 & 15 & 0\\
    Generous Interfaces for Digital Collections~\cite{whitelaw_generous_2015} & Whitelaw & DHQ & 2015 & 16 & 0\\
    Visualizing Cultural Heritage Collections~\cite{windhager_visualization_2018} & Windhager & TVCG & 2018 & 23 & 10\\
    \bottomrule
  \end{tabular}
\label{tab:seed}
\end{table*}

\subsection{Sampling Process}


From the seed sample described above, a semi-automatic snowball article search of references was performed in Scopus in January 2024 both forward (i.e., retrieving all works citing at least one article in the seed) and backward (i.e., all articles cited by at least one article in the seed). From this list, we kept only peer-reviewed conference and journal papers, and thus excluding other types of resources such as web pages, blog posts, or books. For each of the added articles, the provenance was kept as a reference across all the generated intermediate files, including the seed article and whether it was captured either in the forward or backward passes. The snowball search resulted in 1246 unique articles that were retrieved in January 2025. Additionally, we also included 71 papers from all eight editions of the Workshop on Visualization for the Digital Humanities (VIS4DH) \footnote{\url{https://vis4dh.dbvis.de/}}, for a total of 1,317 works that were screened in a subsequent step. Notice that our focus on full-length, peer-reviewed papers, while necessary to ensure sufficient methodological detail for our analysis, means this survey does not capture potentially innovative evaluation practices presented in other formats such as books, posters, or short workshop papers.

\subsection{Screening} 
After eliminating duplicate articles, all entries on the list were assigned to two authors for review. Specifically, we aimed for visualization design studies intended to support research in the humanities (i.e., each selected article presents a development and/or implementation of a visualization solution), which were written in English, published in peer-reviewed journals or conference proceedings since 2013, and included an evaluation of the proposed visualization solution in accordance with the guidelines presented in \cite{isenberg_systematic_2013}. When multiple articles related to the same project were found, we selected the one with the most thorough evaluation. To ensure the reliability and depth of the evaluation evidence, we excluded works that provided too few details on the evaluation procedure, such as those that merely mention user feedback or anecdotal observations in an informal manner. We also excluded articles too short to describe the evaluation in sufficient detail, setting a minimum length of four pages in double-column format. Accordingly, we excluded abstracts, extended abstracts, posters, and other short-form publications that typically do not offer comprehensive methodological accounts. Finally, we excluded works that were not explicitly aimed at supporting humanistic reasoning---such as visualizations of NLP methods intended primarily for evaluating algorithms or computational techniques, without engaging with or enabling domain-specific interpretation in the humanities. 

In this stage, we primarily examined titles and abstracts to assess whether the inclusion criteria were likely to be met; in cases of uncertainty, we briefly scanned the full text to make a more informed decision. Each author independently assessed the assigned articles, followed by one-on-one meetings to resolve discrepancies. In total, 1103 works were excluded in this stage.

\subsection{Eligibility}
\begin{figure}[h]
    \centering
    \includegraphics[width=\linewidth]{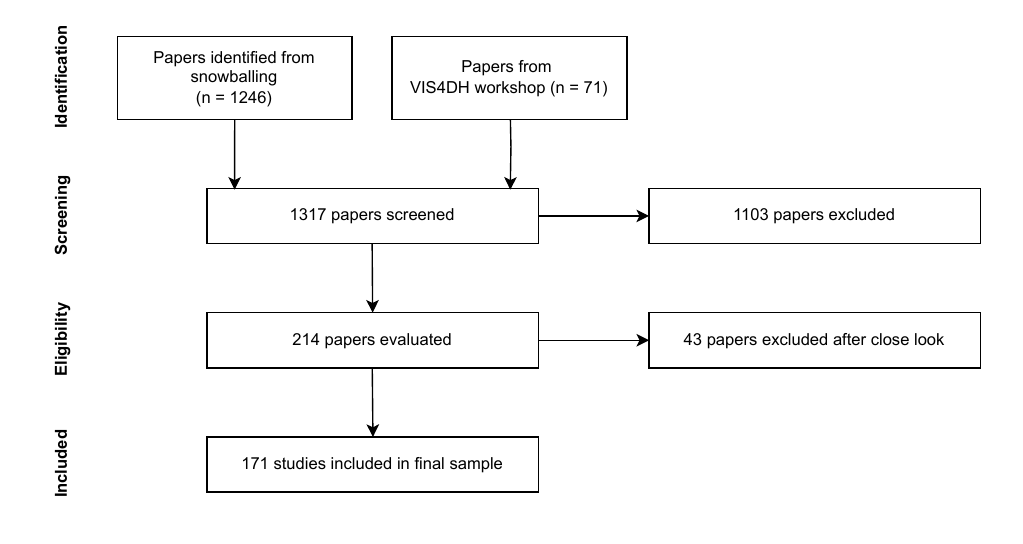}
    \caption{PRISMA flow diagram detailing the literature selection process. Starting from a seed corpus of seminal works, 1,317 records were identified via forward and backward snowballing. The multistage screening process, which filtered for sufficient evaluation detail and relevance to humanities inquiry, resulted in a final corpus of 171 design studies included in the survey.}
    \Description{This PRISMA-style flow diagram summarizes the sampling process for the survey. The search identified 1,246 papers through forward and backward snowballing and 71 papers from the VIS4DH workshop, for a total of 1,317. After screening, 1,103 papers were excluded, leaving 214 for detailed evaluation. Following close review, 43 more were excluded, resulting in 171 studies that were included in the final sample. The figure illustrates the stepwise filtering from identification through screening, eligibility, and inclusion.}
    \label{fig:visxdh_sampling}
\end{figure}

The remaining 214 articles were randomly distributed among the authors of this paper, with each article again assigned to two authors for close review. The assignment process was balanced to ensure that each possible pair of authors reviewed an equal number of articles. Each author independently assessed all the survey categories (discussed in the next section) and re-verified that the article met all inclusion criteria and provided sufficient detail for categorical analysis. Then, bilateral meetings were scheduled again for conflict resolution, followed by a general meeting where non-conclusive categorizations and potential exclusions were discussed and resolved among all authors. In this eligibility stage, the sample was further refined by removing another 43 articles due to missing or insufficient evaluation details, lack of alignment with the focus on supporting humanistic research, or failure to meet length and publication criteria, leaving a final sample of 171 VIS*H design studies, which represents the main dataset analyzed in our survey. A visual summary of the sampling procedure is presented in the PRISMA~\cite{PRISMA} diagram of \autoref{fig:visxdh_sampling}.

\subsection{Publications Sample}
The final sample ($N=171$) is fairly balanced in terms of publication type, comprising 98 journal articles and 73 conference papers. To identify the research areas to which these venues belong, we categorized them manually according to our own research experience. As shown in \autoref{fig:overview} (top left), the sample covers a wide range of publication sources, reflecting the distributed and interdisciplinary nature of the VIS*H field.


\section{Categorization}
\label{sec:categorization}
To capture how standard visualization evaluation practices are being applied in VIS*H design studies, we developed a structured categorization that was used to encode the papers in our sample. The categorization was developed in a two-stage process, first following a ``top-down'' approach, where an initial set of dimensions was selected based on our own expertise during a series of brainstorming sessions. Once this first set was stabilized, in a second step, we followed a ``bottom-up'' approach where we tested the categories with two papers\footnote{The two papers used for this ``bottom-up'' calibration were deliberately selected from markedly different publication venues and traditions: \cite{mccurdy_poemage_2016} appeared in a visualization-focused venue (IEEE TVCG), while \cite{bench_katherine_2020} was published in an arts \& humanities journal (Theater Survey). This contrast was intended to stress-test our preliminary categorization scheme against both the stylistic conventions of the VIS community (with its emphasis on system contributions and technical evaluation) and those of the DH community (with its focus on interpretive framing and critical discourse). By verifying that our categories could be meaningfully applied across such heterogeneous contexts, we ensured that the final scheme would not privilege one publication culture over the other, but instead remain flexible and robust enough to accommodate the diverse styles of reporting found in our sample corpus.} from our sample (\cite{mccurdy_poemage_2016} and \cite{bench_katherine_2020}), giving us the chance to refine the specific components of each category. All authors performed an initial assessment of the categories in these two articles, and individual results and impressions were discussed during multiple joint meetings in which the analytical categories for the main evaluation were jointly refined and finalized. As presented in the following sections, this framework allowed us to analyze the characteristics of visualization design studies in the DH and to identify trends and gaps in evaluation practices. Finally, the whole set of 171 papers in our sample was encoded using the final list of categories. A comprehensive list of the identified dimensions is presented below.

\begin{itemize}
\item \textbf{Publication Source}. To better understand the origin of our sample, we coded to which disciplines the publication sources are associated: arts and humanities (e.g., Historical Methods), digital humanities (e.g., Digital Humanities Quarterly), the Vis4DH workshop series, information sciences (e.g., Information Design Journal), visualization (e.g., TVCG), human factors (like CHI), Geosciences, and interdisciplinary / other (e.g., PLOS ONE).

\item \textbf{Humanities Field}. To map out the diversity and distribution of disciplinary working areas within VIS*H, we coded the specific humanities field for which---or within which---a data visualization has been developed. The examples were coded openly and aggregated bottom-up into seven categories: literature studies and linguistics (including stylometry, translation, and poetry), history, visual cultural heritage (including art history, archaeology, and GLAM--referring to galleries, libraries, archives, and museums), media studies / performative arts (including theater studies, film studies, and musicology), gender studies, non-humanities fields (like geography, environmental studies, science studies), and open tools without a specific focus. Some visualization designs have been assigned to multiple fields.

\item \textbf{Tasks}. To categorize tasks in a DH-specific manner, we used the first level of the TaDiRAH-taxonomy~\cite{borek2021_tadirah} and identified those tasks, which are supported by the visualization: analyzing, capturing, creating, disseminating, enriching, interpreting, or storing data about a humanities topic. 

\item \textbf{Visualization}. To contextualize the evaluations, we also analyzed three key aspects of the visualizations themselves:
\begin{itemize}
    \item \textbf{Multiple Views}. This category identifies the number of views used in a given visualization and examines the complexity of the visualization designed. 
    Given the multifacetted nature of humanities data sources, visualizations with multiple views can better express the interpretative complexity inherent to humanities scholarship \cite{drucker_humanities_2011}.

    \item \textbf{Uncertainty}. Ambiguity and interpretive flexibility are central themes in VIS*H scholarship~\cite{drucker_humanities_2011, janicke_visualizing_2013, coles_slippage_2014, panagiotidou_communicating_2023, benito-santos_data-driven_2020}. This category examines whether uncertainty is addressed in a paper by explicitly discussing or integrating methods for representing uncertainty. 
    
    \item \textbf{Availability}. This category assesses the accessibility of the visualization tool or code developed in each study (e.g., web interface, GitHub). Accessibility of tools and code was included as a category because sustainability and reproducibility are key concerns in both VIS and DH. Coding availability provided us a way to identify whether visualizations remain accessible after publication and whether they function as enduring research artifacts or one-off prototypes.
\end{itemize}



\item \textbf{Study Design}. We distinguish two different forms of study designs: formative evaluations, which aim to inform the visualization design (mostly before and during development), and summative evaluations, aiming to provide evidence on how well a visualization accomplishes its objectives at the end of its development~\cite{summative}. 

\item \textbf{Evaluation Type}. This category classifies the guiding methodological approaches of an evaluation as either quantitative (e.g., time-on-task, number of interactions, structured questionnaires), qualitative (e.g., observation, focus group, interviews), or both (mixed evaluation). 
 


\item \textbf{Evaluation Methods}. To enable a systematic comparison across the diverse corpus of studies surveyed, we normalized the reported evaluation methods into a set of nine high-level categories commonly employed in HCI and visualization research. This category thus provides the backbone of our comparative analysis (see sections \ref{sec:ordinal_model} and \ref{sec:clustering} ) and highlights which techniques are more or less common in VIS*H evaluation. These categories were derived inductively by clustering semantically and methodologically similar activities and were inspired by previous work \cite{Lin2024, xing_review_2025}. The final taxonomy includes: (1) \textbf{Case Study}, encompassing case studies, usage scenarios, close collaboration examples, and walkthroughs grounded in specific data contexts; (2) \textbf{Interviews}, which subsume structured, semi-structured, and unstructured interviews, as well as informal user or domain expert feedback; (3) \textbf{Observation}, including participant observation, video or audio recordings, and pair analytics; (4) \textbf{Think-Aloud} and other verbal protocols, capturing both classical think-aloud and other forms of verbal data elicited during task execution; (5) \textbf{Questionnaires/Surveys}, including standardized instruments such as SUS and TLX, as well as custom Likert-scale surveys; (6) \textbf{Workshops/Co-Design}, which aggregates co-design activities, prototyping sessions, brainstorming meetings, and focus groups; (7) \textbf{Log Analysis/Analytics}, covering web usage logs, tracking of interaction patterns, and movement analytics; (8) \textbf{Inspection-Based Methods}, including heuristic evaluations, cognitive walkthroughs, and other forms of evaluative system inspection; and (9) \textbf{Testing}, comprising usability testing, task completion studies, and performance-driven assessments. 

\item \textbf{Case Study Origin}. Here we distinguish between different origins of case studies or tool demonstrations based on the participating actors, i.e., case studies driven by humanities experts, by data visualization experts, or conducted in close collaboration between both.

\item \textbf{Study Participants}. The \textbf{number} and profile of participants were coded to assess the scale and representativeness of evaluations. This information is indispensable for understanding the robustness of findings and for situating the strength of evidence in DH evaluation studies.  

\item \textbf{Target Users}. Understanding the intended audience of a visualization is essential because evaluation goals and methods vary depending on whether a tool is aimed at domain experts, students, or the general public. This category discloses whether a visualization targets audiences comprised of DH experts (e.g., researchers, GLAM practitioners, teachers), students, or the general public. Sometimes, multiple target groups have been named. 

\item \textbf{Evaluation Quality and Rigor}. To analyze patterns in our corpus, we developed a heuristic for assessing the methodological rigor of each study's evaluation. This approach relies on a holistic, expert judgment to arrive at a consensus score. While not based on a rigid rubric, our assessments were consistently guided by several key factors central to peer review: the clarity of goals, the appropriateness of methods, the depth of analysis, and the degree of critical self-reflection. The scoring process was operationalized as follows: initially, each of the two authors evaluating a particular paper assigned a score based on a 3-point scale (low, moderate, high). Authors were also instructed to document their comments on assessments they found challenging (e.g., ``I am not sure if the evaluation of this paper should be graded as `low' or `moderate, because ...' '')\footnote{Inter-rater agreements at this stage were .623 (Krippendorf's Alpha), as frequently ratings in-between low and moderate or moderate and high were assigned.}. To take into account in-between ratings and to reconcile discrepancies, intermediate levels were introduced when necessary to achieve better alignment of differing evaluations and to provide a more nuanced perspective of the authors' judgments. For example, whenever initial scores of `low' and `moderate' were given, the final negotiated score was typically 2 (low-to-moderate). Also, papers found to contain acceptable but very borderline evaluations were assigned a score of 0. This process resulted in the final 6-point scale used in our analysis, with levels of (0) absolute minimum
, (1) low
, (2) low to moderate
, (3) moderate
, (4) moderate to high
, and (5) high
. With an increasing score, studies typically were better grounded in humanities practices (e.g. more diverse or representative datasets and case studies, experts as study participants, more intense collaboration with humanities researchers), triangulated different methods, applied them more systematically (e.g. reflection of methods), included multiple evaluation cycles, and document their methods according to established quality standards (i.e., providing sufficient details on study participants, procedure, methods, and analysis). While an informal evaluation without any details would be coded as \textit{absolute minimum}, \textit{high} quality evaluations provide all details necessary to understand the procedure and follow the conclusions, are well-grounded in humanities practices, and triangulate multiple methods in different phases of the design process. However, it is critical to state that this score is an inherently subjective construct, not an objective measure of a paper's intrinsic scholarly worth. It reflects our author team's consensus judgment, which is inevitably shaped by the evaluative norms common in top-tier HCI and visualization venues (see Section \ref{sec:positionality}). Consequently, a different team of reviewers with a stronger grounding in interpretive humanities traditions might have assessed the papers differently, producing different results. Therefore, this score should be understood primarily as an \textbf{analytical tool} used within this survey to explore correlations between reported evaluation practices and their perceived rigor from a human-centered computing perspective. Therefore, the subsequent statistical analyses serve an important function: to explore the structure of this score by identifying which specific evaluation techniques and procedures systematically align with it, thereby helping to explain and validate what constitutes our expert team's consensus on evaluation quality.


\item \textbf{Evaluation Challenges}. In an initial assessment of the corpus, we identified several studies explicitly describing methodological challenges that occurred during the evaluation or its conceptualization - or document ``meta-level challenges'' thereafter (including assessments on the overall relevance of visualizations in the context of their core practices). We coded whether such challenges have been described and analyzed them qualitatively. 


\end{itemize}

\section{Coding Results}
\label{sec:summary_results}

\subsection{Overview of Coding Results}

\begin{figure*}
    \centering
    \includegraphics[origin=c,width=0.9\linewidth]{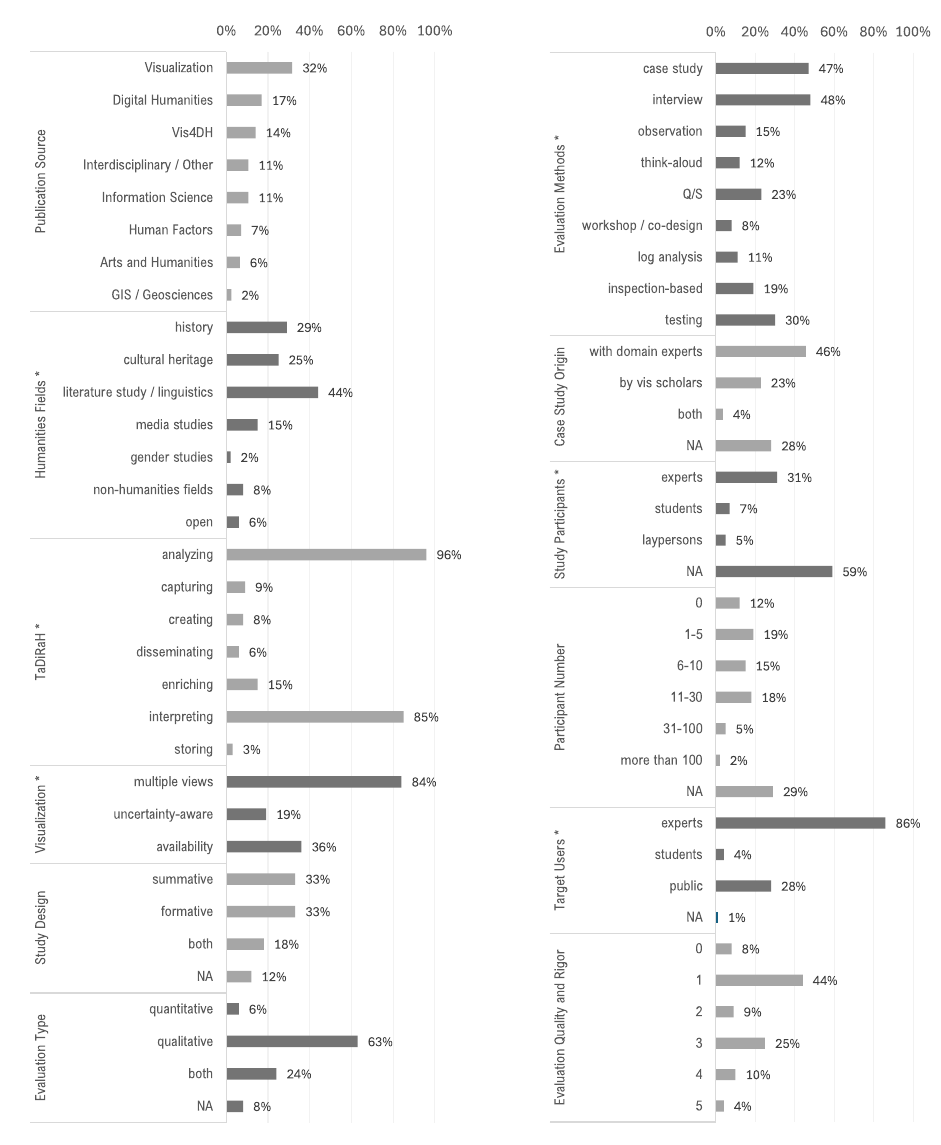}
    \caption{Descriptive overview of the 171 surveyed papers across key dimensions. The data highlights a predominance of Literature and History applications (top left) and a strong focus on Expert users (bottom left). Notably, the distribution of Assessed Rigor Scores (bottom right) is right-skewed, indicating that while low-to-moderate rigor evaluations are common, high-rigor studies remain a minority practice. The * denotes categories where multiple tags can apply.}
    \label{fig:overview}
    \Description{This figure summarizes the coding results from 171 visualization design studies in the humanities. Bar charts show distributions across categories such as publication venues, humanities fields, target users, supported tasks, visualization features, evaluation methods, study designs, participant types and numbers, and evaluation quality. Literature and linguistics dominate the humanities fields, with experts being the primary target users. Most visualizations support analysis and interpretation tasks. Qualitative methods are most common, but mixed-method approaches and larger participant groups tend to yield higher evaluation quality. Overall, the figure highlights uneven distributions, a strong emphasis on expert-oriented tools, and the prevalence of low-to-moderate evaluation rigor.
    }

\end{figure*}

Figure \ref{fig:overview} provides a visual overview of the coding results \footnote{All data supporting this survey are available as supplements. This includes spreadsheets detailing the paper selection at each stage of the PRISMA diagram (\autoref{fig:visxdh_sampling}), as well as the full coding results for each of the 171 included studies.}. In the figure, it can be observed that studies are not evenly distributed across \textbf{humanities fields} but have a major focus within literature studies/linguistics (44\%), followed by history (29\%) and cultural heritage (25\%). Fewer visualization designs have been proposed for media studies/performative arts (15\%), and gender studies (2\%). Some papers also propose open tools (6\%) or non-humanities tools, which can be applied to a humanities topic (8\%).

Visualizations within the DH frequently denote experts (87\%) as their \textbf{target users}, followed by the public (28\%) and students (4\%). The target users differ between the different humanities fields, with most applications for the public developed in the cultural heritage and history fields.
When looking at how the target groups relate to the actual \textbf{study participants}, expert visualizations are actually tested with experts in 31\% of the studies, in 5\% with students, and in 3\% with laypersons (61\% not specified). Public visualizations are tested with laypeople only in 13\% of the studies, more often with experts (26\%), and sometimes with students (9\%, 53\% not specified). Student visualizations are tested with students only in 22\% of the studies, but again more often with experts (56\%, 22\% not specified). In addition to the high number of studies that do not specify their target group at all, the data show a tendency toward expert evaluation, independent of the actual target group. The frequently found usage of students as study participants, is not common in a VIS*H context. There are also interesting counter-intuitive combinations, like an expert visualization tested with laypersons: For example, the Lexichrome visualization \cite{kim_lexichrome_2020} adds color-highlighting to the text for creative writers, but the underlying word-color-associations were tested in a crowdsourcing study with laypersons. 

With regard to the TaDiRAH \textbf{task} taxonomy, most visualizations are designed to support analysis (96\%) and interpretation (85\%) activities. Only a few studies present visualizations for users who want to enrich (15\%), capture (9\%), create (8\%), disseminate (6\%), or store (3\%) data. For example, in Data Hunches \cite{lin_hunches}, experts can \textit{enrich} a visualization with their expert knowledge using comments, ratings, or sketches. Rekall \cite{bardiot_theatre_2020} enables performative artists to \textit{capture}, annotate, visualize, \textit{store}, and preserve their (media) art installations. In DramatVis \cite{hoque_dramatvis_2022}, writers can \textit{create} a character specification for each entity in a text and visually analyze their text's social setup and plot development. In a cultural heritage setting, curators are supported in exploring narrative relations between a cluster of objects for \textit{dissemination} in exhibitions \cite{li_visual_2020}.

\textbf{Study designs} for evaluations of visualizations within the DH are equally often designed as summative and formative studies (approximately one third each). Sometimes, studies combine formative and summative study designs (18\%). Several papers (mainly those categorized as case studies) do not follow a classical study design but rather illustrate the functionality of a tool with a walkthrough-assessment. We coded these cases with `not applicable' (12\%). 
In the next step, we analyzed how the study design is related to evaluation rigor: on each design category (summative, formative, and their combination), at least one study has been rated as high in terms of quality, potentially indicating that evaluation quality could be independent of the study design. However, when we look at median quality scores, combined studies are clearly rated higher (Md=3) than formative (Md=2) and summative ones (Md=1).
The selection of evaluation methods is contingent upon the study design: Summative studies predominantly employ test- and questionnaire-based evaluation methods in comparison to formative studies. Research employing a combination of summative and formative designs more frequently utilizes interviews, supplemented by observations, think-aloud protocols, workshops, and log analyses; however, such studies are less inclined to adopt a case study approach.

Most papers used qualitative \textbf{study methods} (62\%), several combine them with quantitative methods (24\%), and only a few build entirely on quantitative methods (6\%).\footnote{In the remaining cases, the methods have not been sufficiently specified, or have been coded as not applicable.} Prototypical qualitative evaluation methods are interviews, think-aloud procedures, and workshops. Prototypical quantitative evaluation methods are questionnaire-based ones and logfile analyses. Case studies, observation, inspection-based methods, and testing are applied in qualitative and quantitative studies with similar frequency. 

The selection of study methods is related to \textbf{the number of participants}, with most participants in quantitative studies (Md=25), least in qualitative ones (Md=5), and in between for studies applying both methods (Md=14). However, the three studies that report more than 100 participants contain no purely quantitative ones: Mercun et al. \cite{mercun_presenting_2017} amended a larger quantitative experiment (120 participants) with a smaller qualitative observational study (32 participants). D\"ork et al. \cite{dork_monadic_2014} collected qualitative feedback from 124 website visitors. Perovich et al. \cite{perovich} collected qualitative data (ethnographic information, interviews) from five community events with 2 to 60 participants (107 in total).
Similar to the study design, study methods are related to evaluation rigor: it is lowest in qualitative studies (Md=1, low), intermediate in quantitative studies (Md=1.5), and highest in studies which combine qualitative with quantitative methods (Md=3, moderate). Furthermore, those studies rated as minimum in rigor, are all qualitative ones. Also, qualitative studies where rated moderate to high (4) at maximum. These are indicators that qualitative studies tend to be lower in their evaluation quality; however, most quantitative studies have also been rated as low or moderate; only one purely quantitative study was found to be high in evaluation rigor \cite{schmidt}. As \emph{most rigorous}, we categorized mainly evaluations that \textbf{combine quantitative and qualitative study methods} \cite{mehta_metatation_2017, cantareira_exploring_2023, zhang_visual_2023, mercun_presenting_2017, sterman_interacting_2020, hoque_dramatvis_2022}.

Furthermore, we found that a majority of papers do not critically reflect on or discuss their study methodology; only 20\% report some forms of \textbf{challenges} related to their evaluation. Interestingly, such a reflection goes hand in hand with a more rigorous evaluation (Md=3 when challenges are reported versus Md=1 when no challenges are reported).
These challenges include participant recruitment \cite{perovich, schwan_disclosure, feng_ipoet_2022}, small sample sizes \cite{kimura_visualization_2013, meinecke_explaining_2022}, difficult or distorting study settings (for example, in the lab vs. in a natural environment \cite{meinecke_towards_2022,blumenstein_situated_2021}),
the discussion of study methods \cite{blaise_exercises_2019,yuen_enriching_2023,schneier_bigdiva_2018,mercun_presenting_2017}, the reported need for a mixed methods approach \cite{xu_interactive_2014,yuen_enriching_2023, hinrichs_speculative_2016,schneier_bigdiva_2018,blumenstein_situated_2021}, and the insufficient generalizability of an approach \cite{han_hisva_2022, zhang_cohortva_2022,bornhofen, alharbi_transvis_2022}. 
Other papers discussed more DH-specific challenges, which include:
\begin{itemize} 
    \item the heterogeneity of the target group, which either consists of multiple groups \cite{hinrichs_speculative_2016,mayr2022multiple} with different prior knowledge \cite{miller_corpusvis_2022, cantareira_exploring_2023,chen_hierarchical}, or a very diverse multitude of users, especially in public settings like GLAM institutions~\cite{oberbichler_generous_2021,blumenstein_situated_2021,pergantis_user_2022}, 
    \item cross-sectional study designs, which cannot measure longitudinal effects of actual tool use \cite{baumann_annoxplorer_2020,schneier_bigdiva_2018}, especially as non-visualization experts need some time to get accustomed \cite{zhang_visual_2021,bornhofen}
    \item the missing alignment of visualizations with certain aspects of humanistic reasoning \cite{janicke_visualizing_2017, john_multicloud_2018,miller_corpusvis_2022} and the need to adjust or adapt learned analysis procedures to the visual interface \cite{bornhofen,xu_interactive_2014},
    \item the mismatch of the visualization design with the actual data, which are either larger than expected \cite{han_hisva_2022,janicke_interactive_2016,xu_visual_2013}, contain multiple facets and modalities \cite{geng_shakervis_2015,hinrichs_speculative_2016, mayr2022multiple}, contain uncertain and missing data \cite{chang_muse_2023,meinecke_towards_2022,janicke_interactive_2016}, or are visualized as too isolated from a broader context \cite{yuen_enriching_2023,hinrichs_trading_2015},
    \item artificial case study selections, rather than cases grounded in the domain \cite{meinecke_towards_2022},
    \item insufficient or missing collaboration with humanities experts \cite{janicke_visualizing_2017, ehmel_topography_2021,meinecke_towards_2022} or lack of  expert involvement in the evaluation \cite{panagiotidou_implicit_2021}, 
    \item the lack of alignment with or inclusion of specific humanities theories \cite{chang_muse_2023}
    \item the question of whether machine learning algorithms are comparable to humanistic analyses \cite{cheesman2017multi}, and 
    \item difficulties to bridge the gulf between humanities research and artistic practices \cite{bardiot_theatre_2020}.
\end{itemize}

\subsection{Evaluation Quality Scores}
After a final evaluation score was assigned to each paper in the sample via the protocol described in \autoref{sec:categorization}, we computed several summary statistics of interest on the resulting distribution of scores (see right of \autoref{fig:overview}). The mean score was $1.98 \pm1.33$ on the $0$–$5$ scale, and the median and mode were $1$, with the first and third quartiles at $1$ and $3$, respectively. As shown in \autoref{fig:overview}, the distribution is markedly right-skewed: nearly two-thirds of the papers (104/171, 60\%) received a score of $2$ or below, and the modal score of $1$ alone accounts for roughly $45\%$ of the sample (76 papers). Conversely, only about $14\%$ of studies (24 papers) reached the upper end of the scale (scores $4$ or $5$). Across publication sources, Human Factors and Visualization papers reached the highest median scores (Md=$3.0$), Information Science ranked mid-level (Md=$1.5$), and the remaining fields clustered at the lowest level (Md=$1.0$).

\subsection{Evaluation Methods}
The absolute counts for each type of evaluation found in our sample are presented in \autoref{tab:method_stats}. Interviews and case studies dominate the corpus appearing in 83/171 ($23\%$) and 81/171 ($22\%$) of the papers, respectively, while testing (52/171, $14\%$), questionnaires/surveys (39/171, $11\%$), and inspection-based methods (33/171, $9\%$) are also relatively common. Other methods, such as observation (26/171, $7\%$), think-aloud protocols (20/171, $5\%$), log analysis (18/171, $5\%$), and workshops/co-design (13/171, $4\%$), appear less frequently. 

\begin{table}[h]
\caption{
Summary statistics by evaluation method, including usage frequency, average score, and method diversity (i.e., average number of other methods a method can be seen with in our sample). The table reveals that common methods like interviews and case studies tend to score lower and also show less diversity. Less frequent methods, such as think-aloud and log analysis, achieve higher scores and are also combined with more methods on average.
}
\label{tab:method_stats}
\centering
\begin{tabular}{c|c|c|c}
\toprule
\textbf{Method} & \textbf{Frequency (\%)} & \textbf{Score} & \textbf{Diversity} \\
\hline
Interviews              & 82 (23\%) & 2.53 ± 1.28 & 1.78 ± 1.21 \\
Case Study              & 80 (22\%) & 1.40 ± 1.11 & 0.67 ± 1.10 \\
Testing                 & 52 (14\%) & 2.88 ± 1.17 & 2.27 ± 1.09 \\
Quest./Surveys          & 39 (11\%) & 3.05 ± 1.07 & 2.15 ± 1.06 \\
Inspection-Based        & 33 (9\%)  & 2.27 ± 1.35 & 1.93 ± 1.21 \\
Observation             & 26 (7\%)  & 2.84 ± 1.37 & 2.65 ± 1.06 \\
Think-Aloud             & 20 (5\%)  & 3.40 ± 0.99 & 2.70 ± 1.17 \\
Log Analysis            & 18 (5\%)  & 3.50 ± 0.99 & 2.61 ± 1.06 \\
Worksh./Co-Des.     & 13 (4\%)  & 2.31 ± 1.18 & 1.54 ± 0.97 \\
\bottomrule
\end{tabular}
\end{table}

\begin{figure}[h!]
    \centering
    \includegraphics[width=0.75\linewidth]{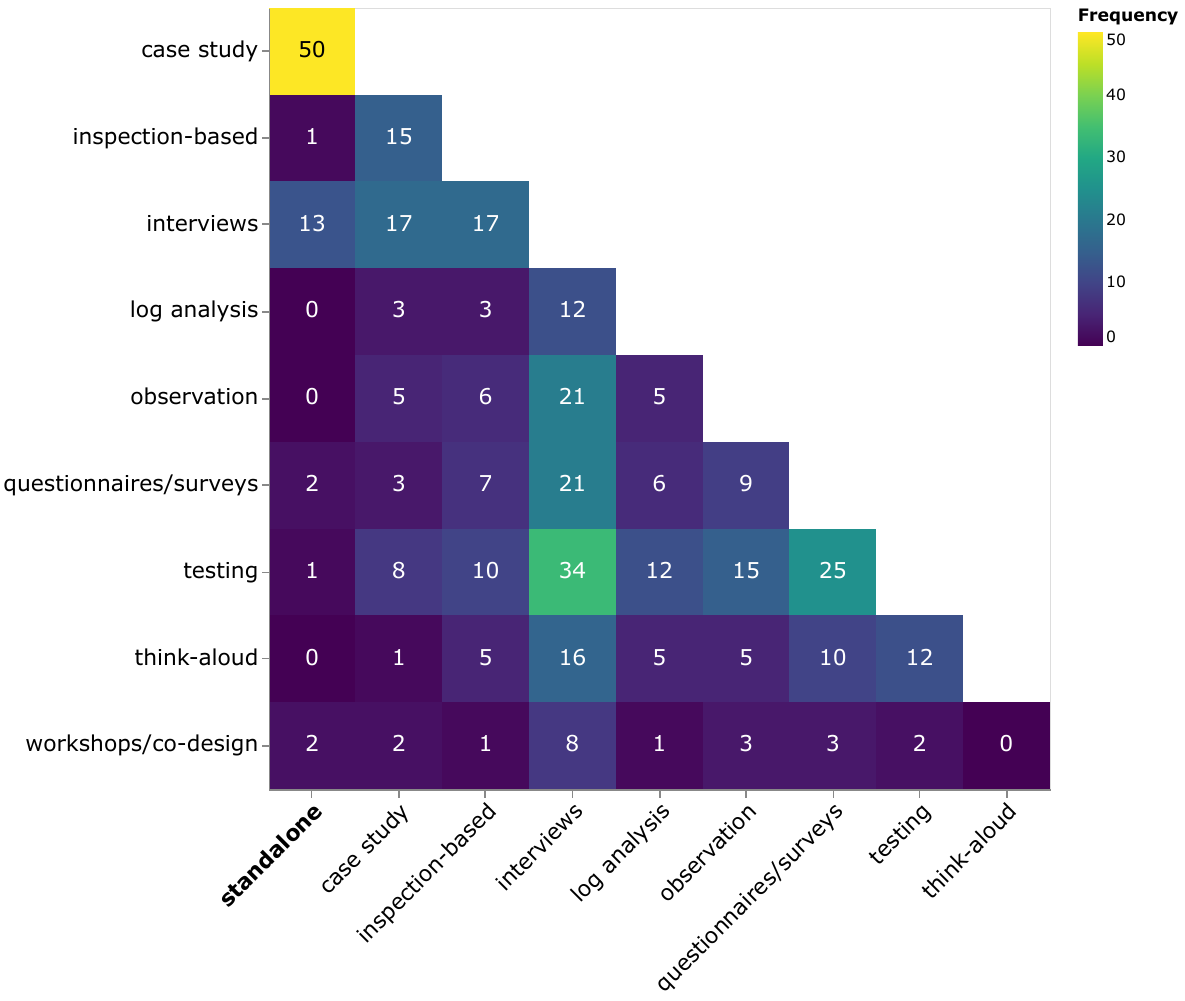}
    \caption{Co-occurrence matrix of evaluation methods found in our sample. Each cell shows the frequency with which two methods were used together in the same study, with brighter colors indicating higher absolute frequencies. Notable patterns include the frequent co-use of interviews with testing (34) and with questionnaires/surveys (21), as well as the predominant use of case studies as standalone methods (50).}
    \label{fig:method_associations_heatmap}
    \Description{This heatmap shows how different evaluation methods co-occur across the surveyed studies. Case studies are the most frequent stand-alone method, appearing alone in 50 papers. Interviews and testing appear widely and often in combination with nearly every other method, with interviews plus testing being the single most common pairing. Questionnaires and surveys are also frequently combined with testing and interviews. In contrast, methods such as log analysis, think-aloud protocols, and workshops or co-design are rarely used alone and appear less frequently overall. The overall pattern highlights a field that relies heavily on case studies and interviews, with more evidence-rich methods integrated less consistently.}
\end{figure}

Additionally, we looked into how evaluation methods are combined in our sample. A first simple approach consisted of counting how many times any two methods appear together. Furthermore, we also captured how many times a method appears alone (i.e., it was the only method employed in the evaluation). The heatmap of \autoref{fig:method_associations_heatmap} shows a very asymmetric pattern in which a handful of methods dominate both as stand-alone approaches and as partners in multiple-method designs. Case studies stand out: 50 papers (out of 81) relied on them alone—by far the largest single count—yet they also co-occur appreciably with inspection-based reviews (15) and interviews (17). Beyond that, interviews and testing act as the main ``connective tissue'' of the matrix: each links to nearly every other method, and their mutual pairing (34) is the most frequent combination, followed closely by testing + questionnaires/surveys (25) and interviews + observation / interviews + questionnaires (both 21). These high counts suggest that researchers often triangulate qualitative insights from interviews with empirical evidence from usability testing or surveys. In contrast, methods such as log analysis, think-aloud, and especially workshops/co-design appear far less often. 

Interestingly, log analysis, think-aloud, and observation-based protocols are the only methods that are never used in isolation. Our data pictures VIS*H as a field that leans heavily on qualitative case-oriented inquiry (63\%), but frequently augments it with quantitative empirical techniques (24\%). 

\section{Analysis of Evaluation Methods and Workflows}
\label{sec:analysis_of_workflows}

The preceding overview mapped the terrain of VIS*H evaluation, revealing a landscape with a clear pattern of preferred techniques, yet also a notable concentration of studies at the lower end of our quality scale. This discrepancy between frequent use and high rigor motivates a deeper investigation into the underlying determinants of what constitutes a robust evaluation and thereby also into the validity of the evaluation quality judgment.

At this point, a central question emerges: \textbf{is superior evaluation quality primarily a function of the number of techniques applied, or is it driven by the substantive nature of the selected methods and their strategic combination?} To address this, we conducted a dual-faceted investigation\footnote{Python and R code to reproduce our analyses is provided in the supplementary materials, visit \url{https://doi.org/10.1145/3772318.3793150} to download them.}. We begin by employing a multivariate regression model to isolate the independent influence of 1) each evaluation method and 2) the number of evaluation methods employed on our quality ratings, allowing us to parse out the contributions of specific techniques from the overall scope of the evaluation protocol. Following this, we apply hierarchical cluster analysis to identify recurrent methodological constellations, revealing the archetypical workflows that define the field's current evaluative paradigms. By integrating these analyses, we aim to construct a more nuanced understanding, moving beyond a simple census of methods to identify the specific practices and workflows that correlate with more insightful evaluation in VIS*H. 
 
\subsection{Relation between Evaluation Methods,  Breadth and Quality Score}
\label{sec:ordinal_model}
The initial analysis presented so far suggests a latent asymmetry in the application of the evaluation methods, potentially indicating an overemphasis on particular instances (e.g., case studies or interviews) to the detriment of other pertinent techniques (e.g., think-aloud or co-design sessions). Also, an important first observation is that certain methods seem to be used in conjunction with more methods on average than others (see diversity scores in \autoref{tab:method_stats} and \autoref{fig:method_associations_heatmap}), which we argued could hypothetically affect the quality score---a question also examined by Lin et al.~\cite[Sec.~5, ``How many evaluations are enough?'']{Lin2024} in their recent review of evaluation practices in information visualization. For example, case studies appear, on average, with significantly fewer other methods than think-aloud protocols (mean diversity score of $0.67$ vs. $2.70$). Beyond that, case studies and interviews seem to be used in isolation more often than other methods. Interestingly, a proportional increase in the scores assigned to papers employing either method can be observed ($1.40\pm1.11$ case studies vs. $3.40\pm0.99$ think-aloud). However, when comparing other methods, such as log analysis (diversity: $2.61\pm1.06$, quality score: $3.50\pm0.99$) and observation (diversity: $2.65\pm1.06$, quality score: $2.85 \pm 1.38$), the relationship between the volume of methods employed in a study and its quality becomes less clear. 

Based on these observations, we sought to answer two specific questions: (1) which methods \textit{independently} predict evaluation quality, if any, and (2) whether the number of methods employed in an evaluation protocol has any effect on perceived quality (Pearson $r = .668$, $p < .01$). We provide the results of our investigation on these two questions hereafter. 

\subsubsection{Modeling rationale and specification}
Inspired by recent work \cite{xing_review_2025}, and to disentangle the influence of \emph{breadth} (i.e., the sheer number of methods used) from the influence of \emph{which} methods are chosen, we modeled the quality scores with a proportional–odds \emph{ordinal} logistic regression. To this end, we fitted the model to the ordered evaluation-quality scores, using the total number of methods plus a set of method-specific dummy variables (with \emph{case study} as the reference category) as predictors.

Our initial proportional–odds ordinal logit model fit on the six-point scale (0–5) showed a significant \emph{omnibus} Brant test, driven by sparse counts at the distribution tails. To respect the ordinal nature of the outcome while satisfying the parallel–slopes assumption, we collapsed the scores into three contiguous, semantically coherent bands: \textbf{low} (0–1), \textbf{medium} (2–3), and \textbf{high} (4–5). Let $Y_i^\star\in\{0,1,2\}$ denote the collapsed score for paper~$i$ (low/medium/high). The model is:

\begin{equation}
\begin{split}
\mathrm{logit}\bigl[\Pr(Y_i^\star > k)\bigr] 
    &= \theta_k - \Bigl(\beta_0\,\textit{method\_count}_i + \sum_{m=1}^{8}\beta_m D_{mi}\Bigr), \\
    &\quad k\in\{0,1\}
\end{split}
\end{equation}

where $\theta_k$ are cut–points, $\textit{method\_count}$ is the total number of distinct methods employed, and $D_{mi}$ are eight binary indicators for the method families.\footnote{\emph{Case study} is omitted and thus serves as the reference.} Parameters were estimated by full maximum likelihood with a logit link using \texttt{MASS::polr} in~R \cite{polr}. Using three levels, the proportional–odds assumption \emph{held}: the Brant test’s omnibus statistic was $\chi^2(9)=15.39$, $p=.081$, and no individual predictor violated parallel slopes (all $p\ge .08$).

\subsubsection{Model fit}
With $N=171$ and nine predictors, the fitted model achieves log-likelihood $-119.01$ (Residual Deviance $=238.03$; $\mathrm{AIC}=260.03$). Coefficients, standard errors, and odds ratios (OR) with 95\% CIs are reported below.

\begin{table}[h]
\caption{Proportional–odds ordinal logistic regression predicting the \emph{collapsed} evaluation quality (low=0, medium=1, high=2). Odds ratios (OR) and 95\% CIs are shown; bold marks p<.05.}
\label{tab:regression}
\centering
\small
\setlength{\tabcolsep}{4pt} 

\begin{tabular}{@{}lccrc@{}}
\toprule
Predictor & $\beta$ & SE & OR [95\% CI] & $p$\\
\midrule
\textbf{Breadth} & & & &\\
\hspace{1em}\#\,methods & 0.337 & 0.455 & 1.40\,[0.57–3.42] & .459\\[2pt]

\textbf{Methods (ref.=Case Study)} & & & &\\
\hspace{1em}Inspection‐based & 0.328 & 0.618 & 1.39\,[0.41–4.66] & .595\\
\hspace{1em}\textbf{Interviews} & \textbf{1.166} & 0.466 & \textbf{3.21}\,[1.29–7.99] & \textbf{.012}\\
\hspace{1em}\textbf{Log analysis} & \textbf{1.582} & 0.675 & \textbf{4.86}\,[1.29–18.28] & \textbf{.019}\\
\hspace{1em}Observation & 0.208 & 0.646 & 1.23\,[0.35–4.37] & .747\\
\hspace{1em}\textbf{Quest./Surveys} & \textbf{1.394} & 0.514 & \textbf{4.03}\,[1.47–11.04] & \textbf{.007}\\
\hspace{1em}Testing & 0.745 & 0.557 & 2.11\,[0.71–6.27] & .181\\
\hspace{1em}Think‐aloud & 0.932 & 0.643 & 2.54\,[0.72–8.96] & .147\\
\hspace{1em}Worksh./Co-Des. & 0.714 & 0.668 & 2.04\,[0.55–7.56] & .285\\
\bottomrule
\end{tabular}
\end{table}

\subsubsection{Key findings}
From the model fit, we extract three key findings: 
\begin{enumerate}
    \item \textbf{Three techniques independently lift quality.}  
            \emph{Log analysis} (OR$\approx 4.86$), \emph{questionnaires/surveys} (OR$\approx 4.03$), and \emph{interviews} (OR$\approx 3.21$) each substantially increase the odds of reaching a higher score relative to a case‐study baseline.
    \item \textbf{Other methods show positive but imprecise effects.}  
          Other evaluation methods, such as \emph{testing} and \emph{think‐aloud}, retain positive coefficients but are not statistically significant at the 95\% significance level. Inspection‐based, observation, and co‐design do not exhibit unique gains once the other methods are controlled. Their positive coefficients might suggest that these approaches contribute to quality only insofar as they are paired with the richer techniques above (this observation is further examined in the next sections \ref{sec:clustering} and \ref{sec:discussion}). 
    \item \textbf{Breadth alone is \emph{not} enough.}  
        After controlling for composition, the total number of methods employed in a study has a small effect on quality ($\beta=0.34$, $p=.46$), implying that the raw correlation between breadth and general evaluation quality seems to be largely compositional rather than additive. This aligns with Lin et al.’s conclusion that there is no formulaic answer to \textit{``how many evaluations are enough''}: the optimal composition is contribution- and goal-dependent, and simply piling on different methods can yield diminishing returns and unsustainable costs~\cite{Lin2024}. As the authors note in their paper, experimental, system, and technique papers tend to benefit from \emph{targeted} mixes (e.g., case studies with quantitative tests; quantitative plus qualitative), chosen to validate the specific claims at hand~\cite[Sec.~5]{Lin2024}, a finding that aligns with our previous observations reported earlier in \autoref{sec:summary_results}. 
\end{enumerate}

\subsection{Archetypical Evaluation Workflows}
\label{sec:clustering}
For the purpose of extracting and analyzing recurrent combinations of evaluation methods in our sample corpus, we conducted a cluster analysis in the following manner: each paper was represented as a binary vector (one-hot encoding) where each dimension signified the presence or absence of a unique evaluation method. We then performed hierarchical clustering using Ward’s minimum variance linkage method on the squared Euclidean distances between these vectors~\cite{ward_hierarchical_1963}, which allowed us to identify distinct, coherent ``archetypes'' of evaluation practices. The binary vector representation is the most direct way to capture the co-occurrence of methods without imposing any artificial ordering or weighting. We chose Ward’s linkage method because its objective is to create compact, well-separated clusters by minimizing the total within-cluster variance at each step of the agglomeration \cite{hastie_elements_2009}. This approach is particularly effective for discovering dense groups of papers that share a highly similar profile of evaluation techniques, making it well-suited to our intended analysis. The squared Euclidean distance is used as it is the metric intrinsically linked to Ward's original formulation, effectively measuring the dissimilarity between papers by counting the number of evaluation methods they do not share in common \cite{everitt_cluster_2011}.

Finally, the optimal number of clusters was identified through empirical evaluation, resulting in the division of the dataset into eight distinct clusters. The summary statistics for each cluster were calculated and are delineated in \autoref{tab:cluster_stats}. The table details, for each cluster, the number of included papers, the number of distinct evaluation methods employed by papers within the cluster (denoted as ``Breadth''), its average diversity (i.e., the average number of methods employed per papers on a given cluster), the most frequently used method, and the mean and median scores of evaluation quality.

\begin{table}[htbp]
\caption{Summary statistics for the eight clusters identified.}
\label{tab:cluster_stats}
\centering
\resizebox{\columnwidth}{!}{%
\begin{tabular}{ccccccccc}
\toprule
Cluster & \# Papers & Breadth & Diversity & Top Method & \multicolumn{2}{c}{Score} \\
\cmidrule(l){6-7}
        &       &      &               &            & Mean (± Std) & Median \\
\midrule
1 &  51 & 1 & 1.00 & Case Study        & 0.92 (± 0.63) & 1.00 \\
2 &  13 & 8 & 2.54 & Co-design         & 2.31 (± 1.18) & 2.00 \\
3 &  17 & 8 & 3.65 & Log Analysis      & 3.53 (± 1.01) & 3.00 \\
4 &  21 & 5 & 2.33 & Inspection-Based  & 1.86 (± 1.06) & 1.00 \\
5 &  18 & 7 & 3.61 & Observation       & 2.83 (± 1.47) & 3.00 \\
6 &  11 & 5 & 3.18 & Think-Aloud       & 3.09 (± 1.04) & 3.00 \\
7 &  13 & 4 & 2.23 & Quest./Surveys               & 2.69 (± 1.18) & 3.00 \\
8 &  27 & 3 & 1.52 & Interviews        & 1.56 (± 0.97) & 1.00 \\
\bottomrule
\end{tabular}%
}
\end{table}

\subsubsection{Clustering Results}
\begin{figure*}[h]
    \centering
    \includegraphics[width=0.75\linewidth]{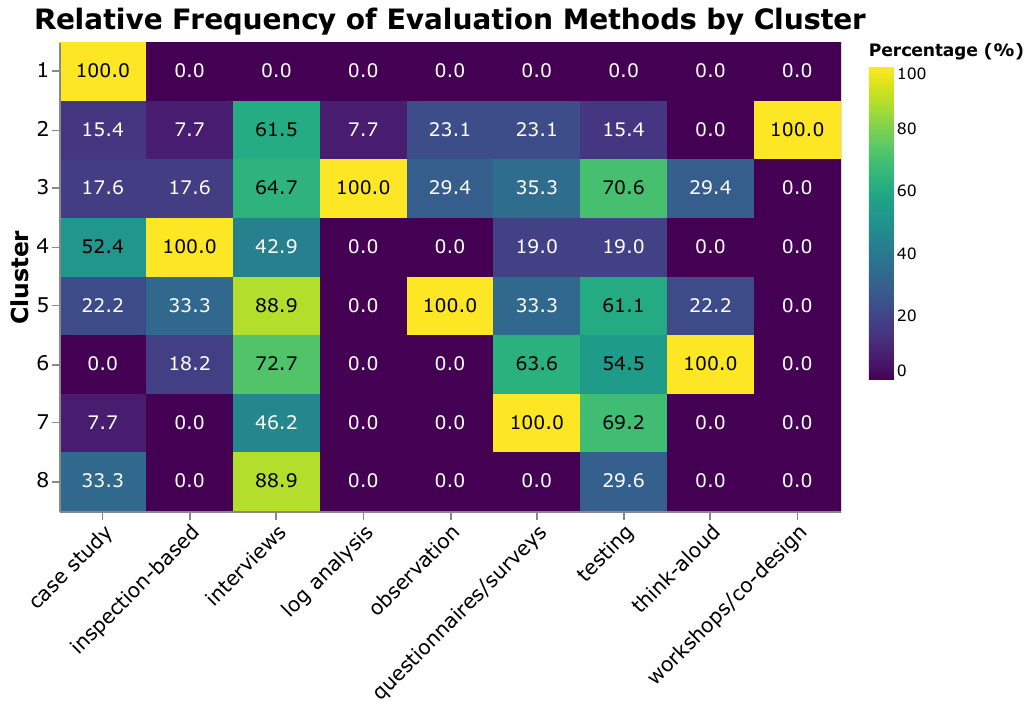}
    \caption{Overview heatmap of the various evaluation methodologies employed across the eight distinct clusters. The x-axis enumerates the nine evaluation methods incorporated in our analysis, while the y-axis represents the cluster number (1-8). Each cell depicts the proportion of papers within each cluster that utilized a specific evaluation method.}
    \Description{This heatmap shows the composition of eight clusters of evaluation workflows, with each row representing a cluster and each column an evaluation method. Each cluster is anchored by one dominant method but differs in how that method is combined with others. Cluster 1 relies exclusively on case studies, usually as standalone evaluations. Cluster 2 centers on workshops and co-design, almost always combined with interviews, surveys, and testing. Cluster 3 is dominated by log analysis and consistently pairs it with multiple complementary methods, including interviews, observation, and testing, making it the most diverse cluster. Cluster 4 is built around inspection-based approaches, sometimes supplemented by interviews or surveys. Cluster 5 emphasizes observation, typically in combination with interviews, surveys, or testing. Cluster 6 consistently uses think-aloud protocols alongside interviews, surveys, and testing. Cluster 7 relies heavily on questionnaires and surveys, often paired with testing or interviews. Cluster 8 is interview-driven, with nearly half the studies using interviews alone, and others combining them with case studies or surveys. Overall, the figure highlights distinct archetypes of evaluation practice, where the strength of each cluster comes not only from its main method but from how it is integrated with others.}
    \label{fig:cluster_composition}
    
\end{figure*}

The heatmap in \autoref{fig:cluster_composition} shows the composition of the identified clusters. The patterns along the y-axis show that certain methods, such as interviews, questionnaires/surveys, and testing, seem to be ubiquitous to all clusters (except for cluster \#1, which captures evaluations consisting only of one or more case studies). Beyond that, each of the eight clusters seems to be dominated by one specific evaluation method that can be found on 100\% of the papers within such cluster (e.g., inspection-based for cluster \#4, observation-based, except for cluster \#8, where 88.9\% of the captured papers relied on interviews). Below, the specific composition of each of these clusters is discussed. 

\begin{figure*}
    \centering
    \includegraphics[origin=c,width=0.75\linewidth]{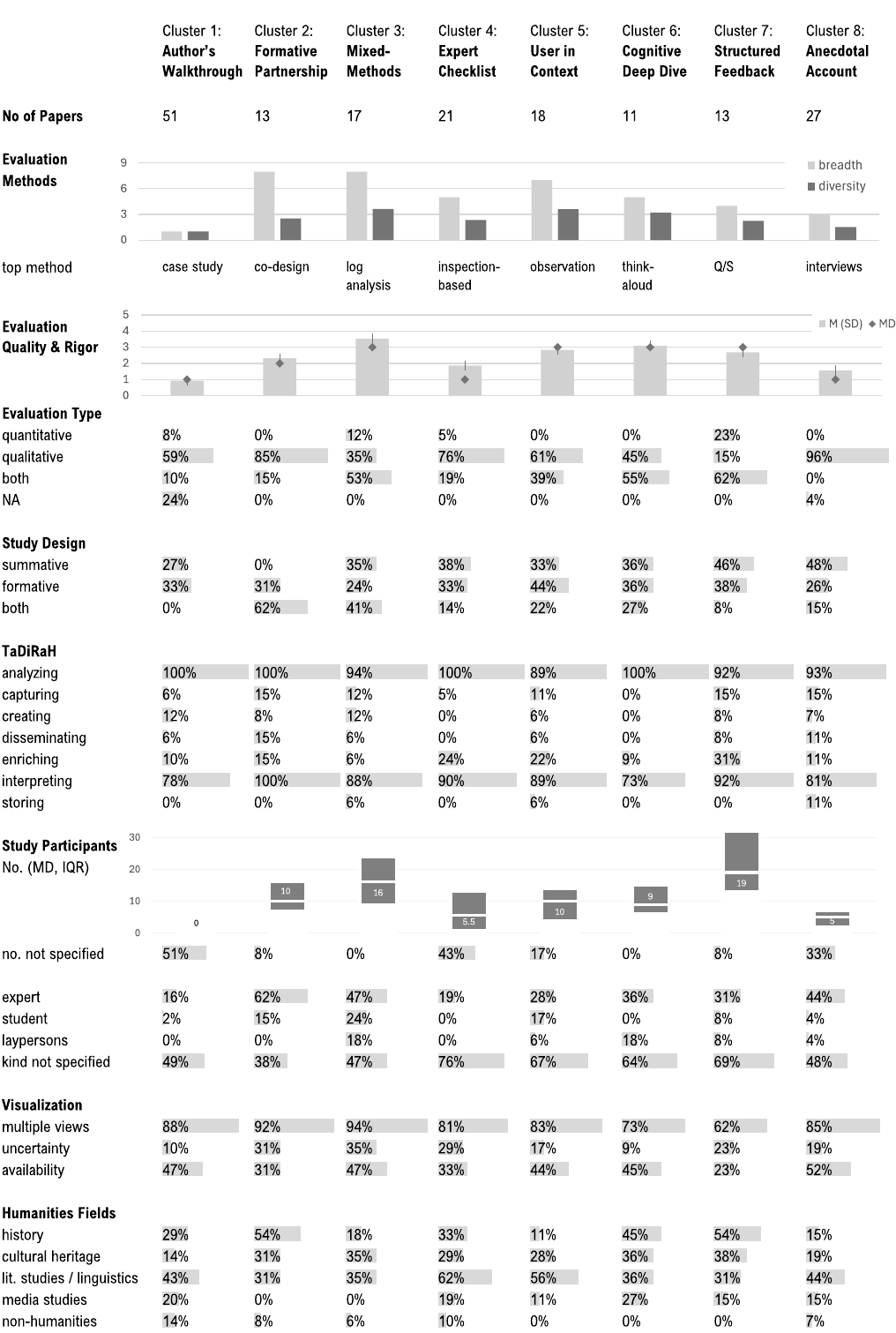}
    \caption{Cluster patterns for different categories: qualitative vs. quantitative methods, summative vs. formative approach, no. (Q1-Q3 range, Md) and kind of study participants, and visualization characteristics (multiple views, uncertainty awareness, availability).}
    \label{fig:cluster_patterns}
    \Description{This figure compares eight evaluation clusters across study methods, design approaches, participant profiles, and visualization features. Cluster 1, based on case studies, relies mostly on qualitative methods with very few or unspecified participants. Cluster 2, anchored in co-design, combines formative and summative approaches and often involves experts. Cluster 3, centered on log analysis, shows the richest mix of qualitative and quantitative methods, larger participant groups, and frequent reporting of uncertainty. Cluster 4, inspection-based, uses small participant numbers and rarely mixes methods. Cluster 5, observation-focused, often combines formative designs with interviews and surveys. Cluster 6, dominated by think-aloud protocols, has the highest share of mixed methods. Cluster 7, built on questionnaires and surveys, involves the largest participant groups. Cluster 8, interview-driven, is strongly qualitative and often lacks detail on participants. Overall, the figure highlights that higher evaluation rigor is linked to clusters with diverse methods, mixed study designs, and well-specified participants.}
\end{figure*}

Taken together, \autoref{fig:cluster_composition} and \autoref{tab:cluster_stats} reveal eight qualitatively distinct archetypical workflows of evaluation practice whose composition is strongly tied to both method diversity and quality score. These are:

\begin{itemize}
    \item \textbf{Cluster~1 – \textit{The Author's Walkthrough}} (51 papers).  
          This cluster illustrates the risk of leaning on a single, largely descriptive method. Papers in this cluster rely exclusively on one or more case studies (breadth\,$=1$, diversity\, $=1.00$) and consequently post the \emph{lowest} quality scores in the corpus ($\bar x=0.92$). Interestingly, 51\% of papers in this cluster do not specify who their participants are, and another 51\% do not specify the kind of participants. With a median of zero specified participants, this strongly suggests that the evaluation is often an author-led demonstration, rather than a user study. Still, some case-study-based evaluations received higher scores than the overall cluster mean (e.g., \cite{benner_sounds_2013,heinicker2023more}). These always followed a formative approach, in which the visualization design is an integral part of the hermeneutic research conducted. Furthermore, this cluster is particularly prevalent in studies on literature and linguistics (43\%), suggesting a potential disciplinary preference for this less rigorous workflow.
    \item \textbf{Cluster~2 – \textit{The Formative Partnership}} (13 papers, best practice example \cite{cantareira_exploring_2023}).  
          Every study here adopts workshops/co-design, typically complemented by ubiquitous methods such as interviews, questionnaires/surveys, and testing (avg.\ 2.54 methods; diversity$=8$). Scores in this cluster are middling ($\bar x=2.31$), suggesting that participatory techniques help but do not, by themselves, guarantee rigorous evaluation. Figure \ref{fig:cluster_patterns} shows that this archetype is characterized by a strong formative approach (62\% use both formative and summative designs) and a high degree of expert involvement (62\% of participants are experts, the highest across all clusters). Its middling score may be explained by its heavy reliance on purely qualitative methods (85\%), often lacking the quantitative validation seen in higher-scoring clusters.
    \item \textbf{Cluster~3 – \textit{The Mixed-Methods Gold Standard}} (17 papers, best practice examples \cite{mehta_metatation_2017,hoque_dramatvis_2022,schmidt}).  
          Dominated by log analysis (100\%), this cluster also exhibits the \emph{highest} method diversity (8 distinct techniques, avg.\ 3.65 per paper), and achieves the best mean score in the dataset ($\bar x=3.53$). This archetype represents a gold standard, consistently combining qualitative and quantitative approaches (53\% use both) and involving a relatively high number of participants (Md=16, all studies reported their number of participants). Here, rich usage analytics appear to be most effective when triangulated with several complementary methods (i.e., interviews, testing, questionnaires/surveys, observation, or think-aloud). Intriguingly, this cluster also contains the highest proportion of design studies that explicitly address uncertainty (35\%), suggesting an association between this workflow and sophistication in visualization design.
    \item \textbf{Cluster~4 – \textit{The Expert Checklist}} (21 papers).  
          Here, heuristic or expert inspections are universal, but they are combined with relatively few additional methods (avg.\ 2.33). Studies within this cluster focus mainly on qualitative inspection (76\%) and involve only a low number of participants (Md=5.5, in 81\% without further specification of who they are). As such, quality remains low ($\bar x=1.86$) and increases only if combined with other methods (e.g. \cite{liu_storyflow_2013}). This archetype is disproportionately found in literary studies (62\%), again pointing to a disciplinary pattern.
    \item \textbf{Cluster~5 – \textit{The User in Context}} (18 papers, best practice examples \cite{mercun_presenting_2017,sterman_interacting_2020}).  
          Observation is present in all papers and paired with a broad mix of interviews, testing, and survey instruments (diversity$=7$, avg.\ 3.61)---often in a formative set-up (66\%). Scores rise to $\bar x=2.83$, indicating that behavioral data boost rigor when embedded in a multi-method design. This cluster shows a healthy mix of qualitative and quantitative methods (39\% use both) and solid participant numbers (Median=10), explaining its strong, above-average performance.
    \item \textbf{Cluster~6 – \textit{The Cognitive Deep Dive}} (11 papers, best practice example \cite{zhang_visual_2023}).  
          Think-aloud protocols anchor this cluster and are coupled with an average of 3.18 other methods. The resulting evaluations are consistently strong ($\bar x=3.09$), underscoring the value of concurrent verbalization for uncovering usability issues, especially when combined with the effective interviews-questionnaires/testing triad. This, along with rigorous participant reporting (Median=8, 0\% not specified), underpins its high quality score. As shown in \autoref{fig:cluster_patterns}, the strength of this archetype lies again in its embrace of mixed methods; it has the highest percentage of studies that combine qualitative and quantitative approaches (55\%).
    \item \textbf{Cluster~7 – \textit{Structured Feedback at Scale}} (13 papers, best practice examples \cite{moretti2016alcide,chen_hierarchical,hu_analyzing_2017,dumas_artvis_2014}).  
          This archetype is a quantitative workhorse, distinguished by having the highest median number of study participants (19) and a strong reliance on mixed-methods (62\% use both qualitative and quantitative). While questionnaires/surveys anchor every study and method breadth is narrower than in top-tier clusters, the resulting scores are consistently respectable ($\bar x=2.69$). This suggests that gathering structured feedback from a larger user group is a robust and reliable evaluation strategy. 
    \item \textbf{Cluster~8 – \textit{The Anecdotal Account}} (27 papers).  
          Interviews are the most common method (88.9\%) in this cluster, which displays the second-lowest diversity score (1.52). In this cluster, almost half the papers (13/27, 48\%) employ interviews as their only evaluation method. This workflow is the most overwhelmingly qualitative of all archetypes (96\%). This methodological narrowness, combined with a low median number of participants (5) and a high rate of non-reporting (33\%), explains its low mean score of 1.56. Like Cluster 1, some papers score higher in rigor \cite{mccurdy_poemage_2016,bares_close_2020,dork_monadic_2014,roessler2019textiles}, which combine interviews with other methods and typically involve a higher number of participants. As in Cluster 1, application domains for studies in this cluster occur predominantly in literary studies and linguistics (43\%)
\end{itemize}

\noindent From the joint analysis of the eight identified clusters, three broad patterns emerge. First, less diverse clusters containing many mono-method evaluations (100\% in cluster 1 and 48\% in cluster 8) score the worst, reinforcing our earlier finding that these minimal evaluation workflows are questionable in VIS*H research. Second, some techniques---log analysis and think-aloud in particular---are consistently embedded within diverse, high-scoring clusters, suggesting a strong association with more rigorous evaluations when used alongside other methods. While this may indicate that such techniques can function as force-multipliers, an alternative explanation is that they are predominantly applied by researchers with strong methodological training, who also design more complex evaluations overall. Third, purely expert-driven approaches (cluster~4) with low participant numbers (or even unspecified ones) fare little better than \textit{Author's Walkthroughs} (cluster 1) and \textit{Anecdotal Accounts} (cluster 8), implying that empirical engagement with users remains crucial. Overall, the cluster analysis highlights the importance of both \emph{selecting} empirically rich methods and \emph{combining} them thoughtfully to attain high-quality evaluations in VIS*H research.

\section{Discussion}
\label{sec:discussion}
Our survey provides a comprehensive empirical snapshot of evaluation practices at the intersection of visualization and the humanities. The results paint a picture of a field with established conventions but also significant room for methodological maturation. While a wide range of evaluation techniques are employed, their application is uneven, and the overall rigor often falls short, with nearly two-thirds of the papers in our sample receiving a low-to-moderate quality score. This section synthesizes the findings from our descriptive overview and statistical analyses to interpret the dominant patterns of evaluation in VIS*H, derive evidence-based recommendations, and connect current practices to the deeper challenges facing the field, for which we provide a deeper discussion in \autoref{sec:challenges}.

\subsection{The Methodological Comfort Zone}
The data reveal a clear ``methodological comfort zone'' in VIS*H evaluation. Case studies and interviews are the most prevalent methods by a significant margin (\autoref{tab:method_stats}), appearing in over 20\% of all papers, respectively. However, our cluster analysis demonstrates that this popularity does not necessarily translate to rigor. The two lowest-scoring evaluation archetypes were Cluster 1 (\textit{``The Author's Walkthrough''}), with a mean score of 0.92, and Cluster 8 (The interview-centric \textit{``The Anecdotal Account''}), with a mean score of 1.56. Together, the monomethod papers found in these clusters account for more than one-third of our sample (63 papers), and they are characterized by a reliance on a single, primarily qualitative method.

This finding suggests that, while case studies and interviews are essential for tool validation and gathering contextual insights, they are insufficient on their own. A simple walkthrough or an unstructured conversation often serves more to demonstrate a tool's features than to rigorously validate its utility or its impact on the scholarly process. In our opinion, this over-reliance on monomethod approaches represents one of the greatest weaknesses in current VIS*H evaluation practice---indicating that common practice is not yet best practice. 

\subsection{Beyond Raw Scores: Disentangling the Drivers of Quality}
A more nuanced story emerges when we compare the simple average scores of methods \autoref{tab:method_stats} with the results of our ordinal logistic regression model (\autoref{tab:regression}). This comparison yielded critical insights: 

\begin{itemize}
    \item \textbf{Evidence-Rich Methods}: The regression model identifies log analysis (OR $\approx 4.86$) and questionnaires/surveys (OR $\approx$ 4.03) as the other most powerful predictors of evaluation quality. This finding suggests that well-designed evaluations are significantly strengthened by the inclusion of methods that capture either direct behavioral evidence (what users do) or structured, quantifiable self-reports (what users think in a systematic way). These ``evidence-rich'' methods provide a crucial check on the purely descriptive and sometimes anecdotal nature of standalone case studies and other simpler evaluation workflows.
    \item \textbf{The Interview Paradox}: Interviews present an interesting case. With a modest average score of 2.53 (\autoref{tab:method_stats}), they might appear to be a mid-tier method. However, the ordinal model reveals that the use of interviews is one of the strongest statistically significant predictors of a higher quality score (OR $\approx 3.21$). This ``paradox'' is explained by the cluster analysis: while many low-rigor studies rely \textit{only} on interviews (cluster 8), high-rigor studies almost always use them as a crucial component to complement and contextualize other forms of user data. The model confirms that it is not the mere presence of interviews, but their integration into a richer, multi-method design that drives quality.
    \item \textbf{The Role of `Force-Multiplier' Methods}: Conversely, other methods obtaining high average quality scores (3.09, \autoref{tab:method_stats}) did not display a proportionally-large effect on the ordinal model. Paired with its high diversity score ($2.70\pm 1.17$), this means these methods --which are never used in isolation-- are indicative of complex, well-designed evaluation workflows that also rely on a variety of other backbone methods (e.g., testing, questionnaires and surveys or interviews).
\end{itemize}

\subsection{The Human Gap: Aligning Evaluation with Humanities Practice} 
\label{sec:humangap}
Our findings also highlight a critical disconnect between evaluation design and the realities of humanistic scholarship. We identified two major gaps:

\begin{enumerate}
    \item \textbf{The Participant Gap}: A critical participant gap threatens the validity of many VIS*H evaluations: while 87\% of tools target experts, only 31\% were evaluated with them. Furthermore, over half of the studies failed to specify their participants at all. Evaluating a tool for literary scholars with a cohort of undergraduate students provides feedback on general usability, not on its capacity to support nuanced, domain-specific research. This gap suggests that many evaluations are testing the tool but not its core value proposition for the intended scholarly community.
    \item \textbf{The Reflexivity Gap}: Only 20\% of papers explicitly discussed the challenges they encountered during their evaluation. Yet, this small cohort of self-reflective papers had a significantly higher median quality score (Md=3) than those that did not (Md=1). This suggests that methodological rigor is not just about flawless execution, but about a critical awareness of the inherent complexities of the task---from participant recruitment to aligning computational tools with interpretive, humanist workflows---and acknowledging these difficulties is a sign of methodological maturity. These empirical gaps are symptoms of the deeper epistemological challenges that we discuss in \autoref{sec:challenges}, such as supporting interpretive complexity and bridging the gulf between computational and humanistic reasoning.
\end{enumerate}

\subsection{Recommendations for VIS*H Evaluators}
Based on this synthesis of our findings, we propose the following evidence-based recommendations for researchers, reviewers, and practitioners in the VIS*H field.

\begin{enumerate}
    \item \textbf{Move Beyond the Monomethod Trap}: Relying solely on a single case study or a set of informal interviews is demonstrably linked to low-rigor evaluations (clusters 1 \& 8). While these methods are valuable starting points, they must be augmented to provide robust validation.
    \item \textbf{Triangulate with Evidence-Rich Methods}: High-quality evaluations stem less from methodological ``check-lists'' and more from the inclusion of complementary techniques that can capture a rich spectrum of user data (e.g., clusters 3, 5, and 6). Researchers seeking rigor should prioritize combining qualitative methods with quantitative or structured evidence. On a limited budget, effective pairings could include:
    \begin{itemize}
        \item \textbf{A think-aloud protocol}, coupled with structured testing sessions, and standardized questionnaires (e.g., SUS, TLX) to link observed behaviors with structured feedback (e.g., \cite{bludau_reading_2020})
        \item \textbf{A logging framework} to capture user behavior in-situ during testing sessions, coupled with follow-up interviews (e.g., \cite{panagiotidou_implicit_2021}). 
    \end{itemize}
    \item \textbf{Prioritize Participant Validity}: The most well-designed evaluation is meaningless if conducted with the wrong participants. We urge researchers to make every effort to recruit from their actual target user group. Furthermore, transparent reporting on participant demographics and expertise should be a non-negotiable standard for publication, as it is fundamental to assessing a study's validity.
    \item \textbf{Embrace and Report on Methodological Challenges}: Our data shows that critical self-reflection typically is a hallmark of papers describing rigorous VIS*H design studies. We recommend that authors dedicate space in their papers to discussing the limitations, trade-offs, and unexpected challenges of their evaluations. This transparency not only strengthens the paper but also contributes valuable knowledge to the entire community about the practical difficulties in VIS*H collaborations.
    \item \textbf{Integrate Evaluation Formatively}: Our results show that studies combining formative and summative methods achieved higher median scores than those using either alone. This supports the established view that evaluation should be a continuous dialogue throughout the design process, not a final validation step. Involving humanities experts early and often in a formative loop is the most effective way to ensure that the final tool is not only usable but genuinely useful and aligned with the practices of humanistic inquiry.
\end{enumerate}




\section{Future Challenges}
\label{sec:challenges}

Where do we expect the field of VIS*H evaluation to go from here? We want to ground our discussion of future challenges in some of our own findings (cf. \autoref{sec:summary_results} on \emph{challenges}) and an increasing number of recommendations for VIS*H design~\cite{hinrichs2017risk, bradley_visualization_2018, lamqaddam_introducing_2020, drucker_visualization_2020, berret_iceberg_2025, windhager2024DH_and_DC}, which emphasize the need for VIS*H to become a real two-way endeavor: A bilateral exchange where data visualization empowers humanities with new methods (VIS$\rightarrow$H, `discovery paradigm'), but equally: where visualization learns from the humanities critical, ethical, and cultural perspectives (VIS$\leftarrow$H, `discursive paradigm') to expand its methodological and discursive repertoire, to ground its representations in multifaceted layers of meaning, and to weave their practice into the broader fabric of human thinking and doing.

Learning from the humanities is demanding, given their theoretical, methodological, and historical complexity \cite{during2020what_were, bianco2012digital}. However, many of their traditions revolve around exploring the \emph{symbolic dimension} of cultural materials and practices---how signs, artifacts, and activities carry meaning and value---but also change and lose them over time, as societies and cultures evolve and transform. The methods developed across fields such as philology, hermeneutics, semiotics, art history, media studies, and historiography differ greatly, yet they converge on \emph{interpretation} as a central activity: connecting symbolic forms to their historical, cultural, and discursive contexts to study and assess their meaning \cite{gadamer_hermeneutik_1990, umberto1994limits}---and to do so \emph{historically and procedurally carefully} (\autoref{subsec:ground_in_provenance}), but also \emph{critically} and \emph{discursively} (\autoref{subsec:ground_in_theory}). In the following, we use the term ``\emph{grounding}'' for summarizing related recommendations with dual resonance. In visualization for the humanities, Lamqaddam and colleagues propose it as a design principle for reducing semantic distance by anchoring representations in contextual dimensions~\cite{lamqaddam_introducing_2020}. In cognitive science, the ``symbol grounding problem'' asks how abstract symbols can acquire meaning, with proposed solutions ranging from perceptual and embodied ties to interactional and socio-cultural accounts of context and use \cite{taddeo2005solving}. Taken together, we adopt grounding as a key concern for future VIS*H design and validation: to connect visualizations to their sources and contexts while at the same time embedding them in traditions of interpretation, critique, and discourse.

\subsection{Grounding Visualizations in Provenance}
\label{subsec:ground_in_provenance}

Through processes of remediation---understood here as representing one medium through another---such as \emph{digitization} and \emph{visualization}, the relationship between cultural objects as sources and scholarly perception and cognition is reconfigured in a non-trivial way. As a result, cultural objects and related information are abstracted into \emph{data} and further transformed into \emph{visual encodings}. These remediation processes are known to enable new perspectives and forms of insight, but also to introduce \emph{digitization losses} and to widen the `semantic distance’ between cultural materials and scholarly representations \cite{lamqaddam_introducing_2020}, while introducing a veneer of scientific objectivity and clarity \cite{drucker_humanities_2011, drucker_visualization_2020}. To counteract this detachment and the illusion of a seamless mediation continuum, VIS*H research has emphasized different forms of grounding that reconnect visualizations to their sources, material qualities, and inherent uncertainties.

\begin{itemize}
 
    \item \textbf{Grounding in Provenance}: Our analysis revealed a significant 'reflexivity gap' in VIS*H evaluation, with only 20\% of papers discussing methodological challenges (\autoref{sec:humangap}). This lack of critical reflection often extends to the data itself, where the journey from cultural source to visual encoding goes unexamined. This empirical gap points to a foundational challenge for the VIS*H field, as across most humanities disciplines, the careful and critical study and evaluation of sources lies at the basis of all further steps of inquiry and interpretation as historically and culturally grounded activities of meaning-making. However, many of the cues needed to assess the origin, authenticity, positionality, originality, relevance, value, or meaning of cultural sources are lost or truncated during digitization procedures~\cite{drucker_humanities_2011, loukissas_taking_2017}. Thus, many design-oriented reflections emphasize the need to amend these digitization losses in the field of VIS*H by embracing any means to trace visualization chains back to their origins first, and to reconstruct subsequent design decisions. This leads to the enhanced \emph{grounding} of abstract representations in \emph{space}, \emph{time}, \emph{cultural context}, and \emph{other topic-related facets} of data provenance through (cont)textual or visual representations of all pertinent transformations \cite{wrisley_pre-visualization_2018, krautli2013known, boyd_davis_can_2021, vancisin2023provenance, windhager2025exploration}. \emph{Future challenges} for provenance-aware evaluation approaches include assessing how well VIS*H projects document and visualize provenance in general \cite{vancisin2023provenance}, the consideration and integration of established provenance frameworks \cite{massari2025representing}, weighing the costs and benefits of added informational complexity~\cite{vancisin2023provenance, windhager2024complexity}, and ensuring that evaluation processes consistently include genuine humanities domain experts alongside VIS practitioners and (digital) humanists (see sec. \ref{sec:summary_results} on the participant gap in VIS*H, as well as \cite{hinrichs2017risk,hall2019design,lamqaddam_introducing_2020}) also with longitudinal and adoption-oriented designs. 
    
    \item \textbf{Grounding in Physicality}: A closely related strategy for minimizing digitization losses in VIS*H is the inclusion of \textit{close reading or viewing} options within ``distant’’ perspectives to accompany diagrammatic views with isomorphic, form-preserving representations of cultural source materials. Beyond the common mantra of providing details on demand, the goal is to ground VIS*H in \emph{physicality}, recreating trust and interpretive intimacy for users who depend on rich representations or direct access to sources \cite{lamqaddam_introducing_2020, windhager_visualization_2018, janicke_visual_2017}. \emph{Future evaluation challenges} for such combined or scalable reading or viewing designs include the development of hybrid non/digital validation frameworks (to assess how tools support cycles of traditional and digital inquiry), for evaluation methods assessing the qualities of isomorphic representations \cite{koller2010research,rybenska2024review}, in methods assessing both fluid or seamful transitions \cite{bludau2025fluidly, hengesbach2022undoing} and a stronger inclusion of domain experts. 

    \item \textbf{Grounding in Uncertainty}: Due to their historical and symbolic nature, humanities data and sources encompass multiple sources of uncertainty. In lieu of quantifiable imprecision, scholars encounter conceptual vagueness, gaps, contradictions, contested interpretations, and the general polysemy of symbolic artifacts. Although humanists are used to handling this uncertainty with their established methods and nuanced terminologies, they are cautious of visualizations that appear too precise or plain and fail to communicate this uncertainty ~\cite{drucker_humanities_2011, drucker_visualization_2020,panagiotidou_communicating_2022}. In response, a growing body of VIS*H reflections calls for \textit{explicit modeling and representation of uncertainty} to raise representation adequacy, transparency, and trust  \cite{panagiotidou_communicating_2022, windhager_visualization_2018, krautli2013known}. \emph{Challenges} for related evaluation endeavors are manifold and include the VIS*H-specific adaptation of accuracy-focused evaluation frameworks and methods \cite{hullman2018pursuit,panagiotidou_communicating_2023}, and also the introduction of the concept of uncertainty early in the design process \cite{benito-santos_data-driven_2020}, together with assessments of tradeoffs regarding added levels of representational complexity \cite{windhager2019uncertainty_of_what}. Given that only 19\% of design study papers in this survey engaged uncertainty at all, this remains an underdeveloped practice that ought to become more central to VIS*H design and assessment.
\end{itemize}

\subsection{Grounding Visualizations in Humanities Theories and Methods}
\label{subsec:ground_in_theory}

If provenance and remediation remind us to be careful about where visualizations come from, theories and methods clarify the intellectual traditions into which they enter. Humanities fields provide frameworks of conceptualization, interpretation, and critique that shape how visualizations are designed, understood, and evaluated—raising new challenges when formal, empiricist criteria intersect with interpretivist, critical, and discursive ones.

\begin{itemize}

    \item \textbf{Grounding in Theory}: Humanities research and practice are guided by theoretical frameworks that shape how questions are posed, phenomena are conceptualized, and findings are interpreted. Theories supply both the elementary terms and distinctions, but also the macroscopic perspectives, which orient activities of inquiry and interpretation \cite{nordberg2023lens, nealon2012theory}. They motivate methodological choices, influence which questions scholars consider worth asking, and establish the quality criteria against which answers and results are judged as meaningful. For VIS*H, engaging with humanities theories and terminologies has been argued to strengthen the conceptual and intellectual footing of their representations, enhancing trust and epistemic alignment \cite{lamqaddam_introducing_2020, windhager2024DH_and_DC}. Selected work in VIS*H documents such a theoretical engagement, for instance, with Actor–Network Theory \cite{latour2012whole}, Post-structuralism \cite{bruggemann2020fold}, or intersectional feminism \cite{dignazio_feminist_2016,akbaba_entanglements_2025}, using theories as productive engines for design and methodological critique. Validating theoretical and terminological grounding efforts, however, raises \emph{evaluation challenges} which signal a larger shift from performance metrics to interpretivist criteria of rigor \cite{meyer2019criteria, berret_iceberg_2025} and require deeper involvement in humanities discourse. Since theories and related design choices cannot be evaluated (i.e., verified or falsified) directly, related reflections need to seek connections to theoretical debates in the humanities or in philosophy of knowledge, where secondary quality criteria such as coherence, fruitfulness, or explanatory power are weighed via discursive validation and contestation \cite{godfrey2009theory}. Addressing evaluation in this way could help close a deficit that has often been noted in the context of DH work \cite{warwick2015building,windhager2024DH_and_DC}.

    \item \textbf{Grounding in Interpretation}: The prevalence of interviews in our dataset (47\% of all papers) indicates a clear desire within VIS*H to engage with the experts’ interpretive processes. However, the low quality scores associated with interview-only evaluations found in the ``Anecdotal Account'' archetype (cluster 8) suggest a gap between intent and execution. This ``interview paradox''—where a valuable method is often implemented superficially—points to a deeper challenge: moving beyond simple feedback to develop workflows that can rigorously capture practices of interpretation. In the humanities, interpretation is understood to pursue and ascribe meaning to cultural materials---texts, images, artworks, performances, or data---by relating them to wider contexts, perspectives, and conceptual frameworks. Successfully executed, it renders implicit structures explicit, condenses what is dispersed, and uncovers connections that are not immediately visible, so that significance can be found where it is not easy to grasp \cite{geertz2017interpretation}. As a basic human activity, interpretation is involved whenever individuals engage with symbolic forms. Yet, the humanities have transformed this everyday practice into methodical forms of inquiry, exemplified by traditions such as hermeneutics and semiotics \cite{gadamer_truth_2004}. Unlike scientific inquiry, which privileges precision, reproducibility, and convergence on well-defined results, interpretive work addresses materials whose semantic openness resists deterministic decoding and closure. Meaning-making in the humanities is therefore cultivated as a collective, non-convergent practice: assembling multiple, position- and context-dependent readings, while still guided by criteria such as evidential grounding, coherence, contextual awareness, depth, and theoretical rigor. This interpretivist stance has recently begun to reshape visualization methodology \cite{meyer2019criteria,berret_iceberg_2025}, which was previously largely subscribed to a positivist orientation, replicable outcomes, and the ideal of convergent and stable knowledge. Interpretivist stances do not dispute the value of this position for many scenarios, but highlight what it misses: the pluralistic, situated, and meaning-pursuing character of visualization as both process and artifact. 
    \emph{Evaluation challenges}: This work is particularly pertinent for VIS*H, where validation choices often intersect directly with interpretive traditions. It casts visualization itself as an interpretive, visual-perceptual practice and calls for evaluation grounded in interpretivist criteria rather than standard scientific metrics. The resulting main challenge (i.e., a potential `loss of rigor') is addressed with the promotion of novel quality criteria---informed, reflexive, abundant, plausible, resonant, and transparent---as revised hallmarks of quality \cite{meyer2019criteria}. As our own analysis indicates, meeting these criteria in practice requires moving beyond the post-hoc reflections often captured by interviews alone. Complementing interviews with methods that capture the interpretive process in situ, such as think-aloud protocols or the analysis of interaction logs during task execution, can provide richer, more direct evidence of scholarly reasoning.

    \item \textbf{Grounding in Critique and Discourse}: We end our reflection on future evaluation challenges in the VIS*H field with a reference to 
    \emph{critique} and \emph{discourse} as interconnected methodological core movements of the humanities. Their role is to agonistically assess and enhance the quality of both interpretations of cultural phenomena and cultural phenomena themselves \cite{liu2012cultural, windhager2025exploration}. Critical-theoretical approaches aim to probe beneath descriptive surface accounts by following a `hermeneutics of suspicion', linked to Marx, Freud, and Nietzsche~\cite{scott2009ricoeur_hermeneutics_suspicion}. Their shared claim is that cultural forms rarely mean what they seem: they operate as visible symptoms of tacit schemes and hidden forces such as economic interests, unconscious drives, or striving for power. Suspicion amplifies interpretation by treating meaning as concealed and distorted, turning interpretation into an act of unveiling deeper forces beneath surface appearances, and of enabling empowerment for those affected by them. For VIS*H, this lens is especially relevant, since visualizations too can naturalize, mask, or distort, and thus call for designs and evaluations that reveal rather than obscure their underlying assumptions \cite{berret_iceberg_2025, dork_critical_2013, hall2022critical, panagiotidou2025critical, windhager2025exploration, akbaba_entanglements_2025}.
    ``Viscursive'' approaches extend and dynamize this agenda by re-embedding visualization in discursive contexts, where individual critical contributions can be evaluated more directly by critical responses, counter arguments, or suggested improvements. They address a core asymmetry: while speech or text supports relatively rapid responses, data visualizations are knowledge and labor-intensive to create and slow to adapt, so visualization discourse often rests and breaks on complex demonstrations, rather than unfolding as continuous debate \cite{he2023enthusiastic}. To counter this, ``viscursive'' strategies call for modes of visual exchange that support interpretive iteration over time---enabling visualization users to participate in evolving conversations and to foster dialogic, critical, and collective meaning-making \cite{walny2020pixelclipper, kauer2024discursive, kauer2025towards}. \emph{Evaluation challenges}: Critical and discursive design strategies reframe evaluation as a continuous, collective process rather than a one-off assessment relegated to design studies. While they require evaluation themselves, their main function is to act as vehicles that return evaluative practice to the humanities community at large, where visual representations can be tested and refined through dialogue, counter-argument, and uptake. Their appraisal will demand a robust mix of empirical and interpretivist criteria. Yet, the emphasis lies on their ability to sustain and facilitate ongoing, discursive refinement of visual representations by the primary actors of humanities research themselves.    

    \end{itemize} 
    
\section{Conclusion}
This survey has revealed a core contradiction in the VIS*H evaluation: the ambition to support deep scholarly inquiry is often paired with methods that only validate surface-level usability. This gap between a tool’s functional performance and its interpretive power defines the field's central challenge. We argue that future evaluation must evolve from a simple validation step to the primary site of negotiation between instrumentalist goals (VIS$\rightarrow$H) and transformative scholarship (VIS$\rightleftarrows$H).

As such, this requires a methodological toolkit that moves beyond assessing utility to engage directly with disciplinary norms, meaning-making practices, and the epistemic potential of visual representation (VIS$\leftarrow$H). Although proposing a comprehensive evaluation framework was beyond the scope of this survey, it represents a clear avenue for future research: the field needs to explicitly define which specific methods are most appropriate for validating the interpretive, contextual, and epistemological aspects that differentiate the humanities from other domains.

To support this ongoing evolution, we have made our full coded dataset available as supplemental material to this paper. We did not explore every research question that could be derived from this corpus, and we invite others to mine these data for further insight. Ultimately, rigorous evaluation in VIS*H demands more than verifying functional utility; it requires assessing how---and whether--—these tools support the core practices that define humanities inquiry. We call on researchers to move beyond their methodological comfort zones and collaboratively forge the critical evaluation practices this ambitious field deserves.

\begin{acks}
This work was initiated by the Dagstuhl Seminar 23381:
``Visualization and the Humanities: Towards a Shared Research Agenda''
\cite{dagstuhl_2024_vis4dh}. The authors want to thank all the colleagues
participating in the initial discussion of evaluation practices and challenges.

\noindent\leavevmode
This work was supported by the \grantsponsor{miciu}{Spanish Ministry of Science, Innovation and Universities}{}
(project ANNOTATE (\grantnum{miciu}{PID2024-156022OB-C31}))
funded by MICIU/\grantsponsor{AEI}{Agencia Estatal de Investigación (AEI)}{https://doi.org/10.13039/501100011033}/\\10.13039/501100011033
and the \grantsponsor{esfplus}{European Social Fund Plus (ESF+)}{}.

\noindent\leavevmode
The work was further funded by the initiative
\grantsponsor{dhinfra}{Digital Humanities Infrastructure in Austria (DHinfra.at)}{}
and by
\grantsponsor{clariah}{CLARIAH-AT}{},
both with support from the Austrian
\grantsponsor{bmfwf}{Austrian Federal Ministry of Education, Science and Research (BMFWF)}{}.

Aida Horaniet Ibañez is funded by the
\grantsponsor{unilu-ias}{University of Luxembourg Institute for Advanced Studies Audacity Program}{}
for the Luxembourg Time Machine.
\end{acks}

\bibliographystyle{ACM-Reference-Format}

\bibliography{references}

@book{gadamer_hermeneutik_1990,
	location = {Tübingen},
	edition = {6},
	title = {Hermeneutik I. Wahrheit und Methode. Grundzüge einer philosophischen Hermeneutik},
	volume = {1},
	series = {Gesammelte Werke},
	volumes = {10},
	publisher = {Mohr},
	author = {Gadamer, Hans-Georg},
	year = {1990},
}

@article{daston_history_2015,
	title = {History of Science and History of Philologies},
	volume = {106},
	issn = {0021-1753, 1545-6994},
	url = {https://www.journals.uchicago.edu/doi/10.1086/681980},
	doi = {10.1086/681980},
	pages = {378--390},
	number = {2},
	journal = {Isis},
	shortjournal = {Isis},
	author = {Daston, Lorraine and Most, Glenn W.},
	urldate = {2024-12-02},
	year = {2015},
	langid = {english},
}

@article{bod_new_2016,
	title = {A New Field: History of Humanities},
	volume = {1},
	issn = {2379-3163, 2379-3171},
	url = {https://www.journals.uchicago.edu/doi/10.1086/685056},
	doi = {10.1086/685056},
	shorttitle = {A New Field},
	pages = {1--8},
	number = {1},
	journal = {History of Humanities},
	shortjournal = {History of Humanities},
	author = {Bod, Rens and Kursell, Julia and Maat, Jaap and Weststeijn, Thijs},
	urldate = {2024-12-19},
	year = {2016},
	langid = {english},
}

@inproceedings{dumas_artvis_2014,
	address = {New York, NY, USA},
	series = {{AVI} '14},
	title = {{ArtVis}: combining advanced visualisation and tangible interaction for the exploration, analysis and browsing of digital artwork collections},
	isbn = {978-1-4503-2775-6},
	shorttitle = {{ArtVis}},
	url = {https://doi.org/10.1145/2598153.2598159},
	doi = {10.1145/2598153.2598159},
	urldate = {2024-09-06},
	booktitle = {Proceedings of the 2014 {International} {Working} {Conference} on {Advanced} {Visual} {Interfaces}},
	publisher = {Association for Computing Machinery},
	author = {Dumas, Bruno and Moerman, Bram and Trullemans, Sandra and Signer, Beat},
	month = may,
	year = {2014},
	pages = {65--72},
}

@inproceedings{dork_monadic_2014,
	address = {New York, NY, USA},
	series = {{CHI} '14},
	title = {Monadic exploration: seeing the whole through its parts},
	isbn = {978-1-4503-2473-1},
	shorttitle = {Monadic exploration},
	url = {https://doi.org/10.1145/2556288.2557083},
	doi = {10.1145/2556288.2557083},
	urldate = {2024-07-05},
	booktitle = {Proceedings of the {SIGCHI} {Conference} on {Human} {Factors} in {Computing} {Systems}},
	publisher = {Association for Computing Machinery},
	author = {Dörk, Marian and Comber, Rob and Dade-Robertson, Martyn},
	month = apr,
	year = {2014},
	note = {tex.ids= dork\_monadic\_2014-1},
	pages = {1535--1544},
}

@article{chang_muse_2023,
	title = {{MUSE}: {Visual} {Analysis} of {Musical} {Semantic} {Sequence}},
	volume = {29},
	copyright = {https://ieeexplore.ieee.org/Xplorehelp/downloads/license-information/IEEE.html},
	issn = {1077-2626, 1941-0506, 2160-9306},
	shorttitle = {{MUSE}},
	url = {https://ieeexplore.ieee.org/document/9781257/},
	doi = {10.1109/TVCG.2022.3175364},
	language = {en},
	number = {9},
	urldate = {2024-07-11},
	journal = {IEEE Transactions on Visualization and Computer Graphics},
	author = {Chang, Baofeng and Sun, Guodao and Li, Tong and Huang, Houchao and Liang, Ronghua},
	month = sep,
	year = {2023},
	pages = {4015--4030},
}

@inproceedings{blumenstein_situated_2021,
	address = {Cham},
	title = {Situated {Visualization} of {Historical} {Timeline} {Data} on {Mobile} {Devices}: {Design} {Study} for a {Museum} {Application}},
	isbn = {978-3-030-85613-7},
	shorttitle = {Situated {Visualization} of {Historical} {Timeline} {Data} on {Mobile} {Devices}},
	doi = {10.1007/978-3-030-85613-7_35},
	language = {en},
	booktitle = {Human-{Computer} {Interaction} – {INTERACT} 2021},
	publisher = {Springer International Publishing},
	author = {Blumenstein, Kerstin and Oliveira, Victor and Boucher, Magdalena and Größbacher, Stefanie and Seidl, Markus and Aigner, Wolfgang},
	editor = {Ardito, Carmelo and Lanzilotti, Rosa and Malizia, Alessio and Petrie, Helen and Piccinno, Antonio and Desolda, Giuseppe and Inkpen, Kori},
	year = {2021},
	pages = {536--557},
}

@inproceedings{li_visual_2020,
	title = {A {Visual} {System} for {Exploring} {Digital} {Cultural} {Relics} and {Generating} {Narrations}},
	url = {https://ieeexplore.ieee.org/abstract/document/9421982?casa_token=lu5zMtR9XCAAAAAA:Scn98-YQH8NNxt21fwpf5OOwh4TxiIOqPdapYw3lZqajOUGImGbsfx5FTnFXaOoQSCCsQSau},
	doi = {10.1109/ITCA52113.2020.00134},
	urldate = {2024-07-05},
	booktitle = {2020 2nd {International} {Conference} on {Information} {Technology} and {Computer} {Application} ({ITCA})},
	author = {Li, Xinxin and Zhang, Jiawan},
	month = dec,
	year = {2020},
	note = {tex.ids= li\_visual\_2020},
	pages = {614--619},
}

@inproceedings{kim_lexichrome_2020,
  ids = {kim_lexichrome_2020-1},
  title = {Lexichrome: {{Text Construction}} and {{Lexical Discovery}} with {{Word-Color Associations Using Interactive Visualization}}},
  shorttitle = {Lexichrome},
  booktitle = {Proceedings of the 2020 {{ACM Designing Interactive Systems Conference}}},
  author = {Kim, Chris and Hinrichs, Uta and Mohammad, Saif M. and Collins, Christopher},
  year = {2020},
  month = jul,
  series = {{{DIS}} '20},
  pages = {477--488},
  publisher = {Association for Computing Machinery},
  address = {New York, NY, USA},
  doi = {10.1145/3357236.3395503},
  urldate = {2025-08-26},
  isbn = {978-1-4503-6974-9}
}

@book{terras_defining_2013,
	address = {London},
	title = {Defining digital humanities: a reader},
	isbn = {978-1-315-57625-1},
	shorttitle = {Defining digital humanities},
	language = {eng},
	publisher = {Routledge},
	editor = {Terras, Melissa M. and Nyhan, Julianne and Vanhoutte, Edward},
	year = {2013},
	note = {OCLC: 952728441},
}

@article{bares_close_2020,
	title = {Close reading for visualization evaluation},
	volume = {40},
	url = {https://ieeexplore.ieee.org/abstract/document/9117084/},
	number = {4},
	urldate = {2024-04-05},
	journal = {IEEE Computer Graphics and Applications},
	author = {Bares, Annie and Keefe, Daniel F. and Samsel, Francesca},
	year = {2020},
	note = {Publisher: IEEE},
	pages = {84--95},
}

@article{bludau_reading_2020,
	title = {Reading {Traces}: {Scalable} {Exploration} in {Elastic} {Visualizations} of {Cultural} {Heritage} {Data}},
	volume = {39},
	issn = {0167-7055, 1467-8659},
	shorttitle = {Reading {Traces}},
	url = {https://onlinelibrary.wiley.com/doi/10.1111/cgf.13964},
	doi = {10.1111/cgf.13964},
	language = {en},
	number = {3},
	urldate = {2024-04-05},
	journal = {Computer Graphics Forum},
	author = {Bludau, M.‐J. and Brüggemann, V. and Busch, A. and Dörk, M.},
	month = jun,
	year = {2020},
	pages = {77--87},
}

@article{panagiotidou_implicit_2021,
	title = {Implicit error, uncertainty and confidence in visualization: an archaeological case study},
	volume = {28},
	shorttitle = {Implicit error, uncertainty and confidence in visualization},
	url = {https://ieeexplore.ieee.org/abstract/document/9451614/},
	number = {12},
	urldate = {2024-04-05},
	journal = {IEEE Transactions on Visualization and Computer Graphics},
	author = {Panagiotidou, Georgia and Vandam, Ralf and Poblome, Jeroen and Moere, Andrew Vande},
	year = {2021},
	note = {Publisher: IEEE},
	pages = {4389--4402},
}

@article{ehmel_topography_2021,
	title = {Topography of {Violence}: {Considerations} for {Ethical} and {Collaborative} {Visualization} {Design}},
	volume = {40},
	issn = {0167-7055, 1467-8659},
	shorttitle = {Topography of {Violence}},
	url = {https://onlinelibrary.wiley.com/doi/10.1111/cgf.14285},
	doi = {10.1111/cgf.14285},
	language = {en},
	number = {3},
	urldate = {2024-04-05},
	journal = {Computer Graphics Forum},
	author = {Ehmel, F. and Brüggemann, V. and Dörk, M.},
	month = jun,
	year = {2021},
	pages = {13--24},
}

@article{xu_visual_2013,
	title = {Visual {Analysis} of {Set} {Relations} in a {Graph}},
	volume = {32},
	issn = {0167-7055, 1467-8659},
	url = {https://onlinelibrary.wiley.com/doi/10.1111/cgf.12093},
	doi = {10.1111/cgf.12093},
	language = {en},
	number = {3pt1},
	urldate = {2024-04-05},
	journal = {Computer Graphics Forum},
	author = {Xu, Panpan and Du, Fan and Cao, Nan and Shi, Conglei and Zhou, Hong and Qu, Huamin},
	month = jun,
	year = {2013},
	pages = {61--70},
}

@article{yuen_enriching_2023,
	title = {Enriching {Museum} {Student} {Users}’ {Curriculum} {Interpretation}, {Learning}, and {Worldview}},
	volume = {17},
	url = {https://search.proquest.com/openview/9ed933e12c4b93fdda227d465e92952f/1?pq-origsite=gscholar&cbl=5531554},
	number = {1},
	urldate = {2024-04-05},
	journal = {The International Journal of the Inclusive Museum},
	author = {Yuen, Yee Lin Elaine and Leslie, Catherine Amoroso},
	year = {2023},
	note = {Publisher: Common Ground Research Networks},
	pages = {23},
}

@article{pergantis_user_2022,
	title = {User evaluation and metrics analysis of a prototype web-based federated search engine for art and cultural heritage},
	volume = {13},
	url = {https://www.mdpi.com/2078-2489/13/6/285},
	number = {6},
	urldate = {2024-04-05},
	journal = {Information},
	author = {Pergantis, Minas and Varlamis, Iraklis and Giannakoulopoulos, Andreas},
	year = {2022},
	note = {Publisher: MDPI},
	pages = {285},
}

@inproceedings{hoque_dramatvis_2022,
	address = {Virtual Event Australia},
	title = {{DramatVis} {Personae}: {Visual} {Text} {Analytics} for {Identifying} {Social} {Biases} in {Creative} {Writing}},
	isbn = {978-1-4503-9358-4},
	shorttitle = {{DramatVis} {Personae}},
	url = {https://dl.acm.org/doi/10.1145/3532106.3533526},
	doi = {10.1145/3532106.3533526},
	language = {en},
	urldate = {2024-04-05},
	booktitle = {Designing {Interactive} {Systems} {Conference}},
	publisher = {ACM},
	author = {Hoque, Md Naimul and Ghai, Bhavya and Elmqvist, Niklas},
	month = jun,
	year = {2022},
	pages = {1260--1276},
}

@incollection{blaise_exercises_2019,
	address = {Cham},
	title = {Exercises in {Unstyling} {Texts}: {Formalisation} and {Visualisation} of a {Narrative}’s [{Space}, {Time}, {Actors}, {Motion}] {Components}},
	volume = {914},
	isbn = {978-3-319-99700-1 978-3-319-99701-8},
	shorttitle = {Exercises in {Unstyling} {Texts}},
	url = {http://link.springer.com/10.1007/978-3-319-99701-8_2},
	language = {en},
	urldate = {2024-04-05},
	booktitle = {Knowledge {Discovery}, {Knowledge} {Engineering} and {Knowledge} {Management}},
	publisher = {Springer International Publishing},
	author = {Blaise, Jean-Yves and Dudek, Iwona},
	editor = {Fred, Ana and Dietz, Jan and Aveiro, David and Liu, Kecheng and Bernardino, Jorge and Filipe, Joaquim},
	year = {2019},
	doi = {10.1007/978-3-319-99701-8_2},
	note = {Series Title: Communications in Computer and Information Science},
	pages = {28--53},
}

@article{xu_interactive_2014,
	title = {Interactive visualization for curatorial analysis of large digital collection},
	volume = {13},
	issn = {1473-8716, 1473-8724},
	url = {http://journals.sagepub.com/doi/10.1177/1473871612473590},
	doi = {10.1177/1473871612473590},
	language = {en},
	number = {2},
	urldate = {2024-03-27},
	journal = {Information Visualization},
	author = {Xu, Weijia and Esteva, Maria and Jain, Suyog D and Jain, Varun},
	month = apr,
	year = {2014},
	pages = {159--183},
}

@article{drucker_performative_2013,
	title = {Performative {Materiality} and {Theoretical} {Approaches} to {Interface}},
	volume = {7},
	copyright = {© 2013. This work is published under http://creativecommons.org/licenses/by-nd/4.0/ (the “License”). Notwithstanding the ProQuest Terms and Conditions, you may use this content in accordance with the terms of the License.},
	url = {https://www.proquest.com/docview/2555208733/abstract/FABF0B3822274F25PQ/1},
	language = {English},
	number = {1},
	urldate = {2024-03-27},
	journal = {Digital Humanities Quarterly},
	author = {Drucker, Johanna},
	year = {2013},
	note = {Place: Providence, Providence
Section: Articles},
}

@article{hinrichs_trading_2015,
	title = {Trading {Consequences}: {A} {Case} {Study} of {Combining} {Text} {Mining} and {Visualization} to {Facilitate} {Document} {Exploration}},
	volume = {30},
	issn = {2055-7671},
	shorttitle = {Trading {Consequences}},
	url = {https://doi.org/10.1093/llc/fqv046},
	doi = {10.1093/llc/fqv046},
	number = {suppl\_1},
	urldate = {2024-03-27},
	journal = {Digital Scholarship in the Humanities},
	author = {Hinrichs, Uta and Alex, Beatrice and Clifford, Jim and Watson, Andrew and Quigley, Aaron and Klein, Ewan and Coates, Colin M.},
	month = dec,
	year = {2015},
	pages = {i50--i75},
}

@article{alvarado_digital_2012,
	title = {The digital humanities situation},
	journal = {Debates in the digital humanities},
	author = {Alvarado, Rafael C.},
	year = {2012},
	pages = {50--55},
}

@article{janicke_visualizing_2013,
	title = {Visualizing uncertainty: {How} to use the fuzzy data of 550 medieval texts},
	volume = {2013},
	journal = {Proceedings of the Digital Humanities},
	author = {Jänicke, S and Wrisley, David Joseph},
	year = {2013},
}

@inproceedings{benner_sounds_2013,
	title = {The {Sounds} of the {Psalter}: {Computational} {Analysis} of {Phonological} {Parallelism} in {Biblical} {Hebrew} {Poetry}.},
	shorttitle = {The {Sounds} of the {Psalter}},
	booktitle = {{DH}},
	author = {Benner, Drayton Callen},
	year = {2013},
	pages = {105--106},
}

@article{coles_slippage_2014,
	title = {Slippage, spillage, pillage, bliss: {Close} reading, uncertainty, and machines},
	shorttitle = {Slippage, spillage, pillage, bliss},
	journal = {Wester Humanities Review},
	author = {Coles, Katharine},
	year = {2014},
	pages = {39--65},
}

@article{loukissas_taking_2017,
	title = {Taking {Big} {Data} apart: local readings of composite media collections},
	volume = {20},
	issn = {1369-118X},
	shorttitle = {Taking {Big} {Data} apart},
	url = {https://doi.org/10.1080/1369118X.2016.1211722},
	doi = {10.1080/1369118X.2016.1211722},
	number = {5},
	urldate = {2018-09-18},
	journal = {Information, Communication \& Society},
	author = {Loukissas, Yanni Alexander},
	month = may,
	year = {2017},
	pages = {651--664},
}

@inproceedings{dork_critical_2013,
	address = {New York, NY, USA},
	series = {{CHI} {EA} '13},
	title = {Critical {InfoVis}: {Exploring} the {Politics} of {Visualization}},
	isbn = {978-1-4503-1952-2},
	shorttitle = {Critical {InfoVis}},
	url = {http://doi.acm.org/10.1145/2468356.2468739},
	doi = {10.1145/2468356.2468739},
	urldate = {2018-09-18},
	booktitle = {{CHI} '13 {Extended} {Abstracts} on {Human} {Factors} in {Computing} {Systems}},
	publisher = {ACM},
	author = {Dörk, Marian and Feng, Patrick and Collins, Christopher and Carpendale, Sheelagh},
	year = {2013},
	pages = {2189--2198},
}

@inproceedings{dignazio_feminist_2016,
	address = {New York, NY, USA},
	series = {{CHI} '16},
	title = {A {Feminist} {HCI} {Approach} to {Designing} {Postpartum} {Technologies}: "{When} {I} {First} {Saw} a {Breast} {Pump} {I} {Was} {Wondering} if {It} {Was} a {Joke}"},
	isbn = {978-1-4503-3362-7},
	shorttitle = {A {Feminist} {HCI} {Approach} to {Designing} {Postpartum} {Technologies}},
	url = {http://doi.acm.org/10.1145/2858036.2858460},
	doi = {10.1145/2858036.2858460},
	urldate = {2018-09-18},
	booktitle = {Proceedings of the 2016 {CHI} {Conference} on {Human} {Factors} in {Computing} {Systems}},
	publisher = {ACM},
	author = {D'Ignazio, Catherine and Hope, Alexis and Michelson, Becky and Churchill, Robyn and Zuckerman, Ethan},
	year = {2016},
	pages = {2612--2622},
}

@article{cantareira_exploring_2023,
	title = {Exploring {Interpersonal} {Relationships} in {Historical} {Voting} {Records}},
	volume = {42},
	issn = {1467-8659},
	url = {https://onlinelibrary.wiley.com/doi/abs/10.1111/cgf.14824},
	doi = {10.1111/cgf.14824},
	language = {en},
	number = {3},
	urldate = {2024-03-27},
	journal = {Computer Graphics Forum},
	author = {Cantareira, G. D. and Xing, Y. and Cole, N. and Borgo, R. and Abdul-Rahman, A.},
	year = {2023},
    pages = {211--221},
}

@article{schneier_bigdiva_2018,
	title = {{BigDIVA} and {Networked} {Browsing}: {A} {Case} for {Generous} {Interfacing} and {Joyous} {Searching}},
	volume = {012},
	issn = {1938-4122},
	shorttitle = {{BigDIVA} and {Networked} {Browsing}},
	number = {2},
	journal = {Digital Humanities Quarterly},
	author = {Schneier, Joel and Stinson, Timothy and Davis, Matthew},
	month = jul,
	year = {2018},
}

@article{oberbichler_generous_2021,
	title = {Generous and inviting interfaces revisited: {Examples} of designing visual structures for digital archives},
	volume = {26},
	copyright = {https://benjamins.com/content/customers/rights},
	issn = {0142-5471, 1569-979X},
	shorttitle = {Generous and inviting interfaces revisited},
	url = {http://www.jbe-platform.com/content/journals/10.1075/idj.20028.obe},
	doi = {10.1075/idj.20028.obe},
	language = {en},
	number = {2},
	urldate = {2024-03-26},
	journal = {Information Design Journal},
	author = {Oberbichler, Sarah and Gallner-Holzmann, Katharina and Hug, Theo},
	month = dec,
	year = {2021},
	pages = {157--174},
}

@article{hu_analyzing_2017,
	title = {Analyzing and visualizing ancient {Maya} hieroglyphics using shape: {From} computer vision to {Digital} {Humanities}},
	volume = {32},
	issn = {2055-7671},
	shorttitle = {Analyzing and visualizing ancient {Maya} hieroglyphics using shape},
	url = {https://doi.org/10.1093/llc/fqx028},
	doi = {10.1093/llc/fqx028},
	number = {suppl\_2},
	urldate = {2024-03-25},
	journal = {Digital Scholarship in the Humanities},
	author = {Hu, Rui and Gayol, Carlos Pallán and Odobez, Jean-Marc and Gatica-Perez, Daniel},
	month = dec,
	year = {2017},
	pages = {ii179--ii194},
}

@article{feng_ipoet_2022,
	title = {{iPoet}: interactive painting poetry creation with visual multimodal analysis},
	volume = {25},
	issn = {1875-8975},
	shorttitle = {{iPoet}},
	url = {https://doi.org/10.1007/s12650-021-00780-0},
	doi = {10.1007/s12650-021-00780-0},
	language = {en},
	number = {3},
	urldate = {2024-03-25},
	journal = {Journal of Visualization},
	author = {Feng, Yingchaojie and Chen, Jiazhou and Huang, Keyu and Wong, Jason K. and Ye, Hui and Zhang, Wei and Zhu, Rongchen and Luo, Xiaonan and Chen, Wei},
	month = jun,
	year = {2022},
	pages = {671--685},
}

@article{zhang_visual_2023,
	title = {Visual {Reasoning} for {Uncertainty} in {Spatio}-{Temporal} {Events} of {Historical} {Figures}},
	volume = {29},
	issn = {1941-0506},
	url = {https://ieeexplore.ieee.org/document/9695348},
	doi = {10.1109/TVCG.2022.3146508},
	number = {6},
	urldate = {2024-03-25},
	journal = {IEEE Transactions on Visualization and Computer Graphics},
	author = {Zhang, Wei and Tan, Siwei and Chen, Siming and Meng, Linghao and Zhang, Tianye and Zhu, Rongchen and Chen, Wei},
	month = jun,
	year = {2023},
	note = {Conference Name: IEEE Transactions on Visualization and Computer Graphics},
	pages = {3009--3023},
}

@article{zhang_visual_2021,
	title = {Visual storytelling of {Song} {Ci} and the poets in the social–cultural context of {Song} dynasty},
	volume = {5},
	issn = {2468-502X},
	url = {https://www.sciencedirect.com/science/article/pii/S2468502X21000607},
	doi = {10.1016/j.visinf.2021.12.002},
	number = {4},
	urldate = {2024-03-25},
	journal = {Visual Informatics},
	author = {Zhang, Wei and Ma, Qian and Pan, Rusheng and Chen, Wei},
	month = dec,
	year = {2021},
	pages = {34--40},
}

@article{mehta_metatation_2017,
	title = {Metatation: {Annotation} as {Implicit} {Interaction} to {Bridge} {Close} and {Distant} {Reading}},
	volume = {24},
	issn = {1073-0516},
	shorttitle = {Metatation},
	url = {https://dl.acm.org/doi/10.1145/3131609},
	doi = {10.1145/3131609},
	number = {5},
	urldate = {2024-03-19},
	journal = {ACM Transactions on Computer-Human Interaction},
	author = {Mehta, Hrim and Bradley, Adam and Hancock, Mark and Collins, Christopher},
	month = nov,
	year = {2017},
	pages = {35:1--35:41},
}

@article{lamqaddam_introducing_2020,
	title = {Introducing layers of meaning ({LoM}): {A} framework to reduce semantic distance of visualization in humanistic research},
	volume = {27},
	shorttitle = {Introducing layers of meaning ({LoM})},
	url = {https://ieeexplore.ieee.org/abstract/document/9222293/},
	number = {2},
	urldate = {2024-02-15},
	journal = {IEEE Transactions on Visualization and Computer Graphics},
	author = {Lamqaddam, Houda and Moere, Andrew Vande and Abeele, Vero Vanden and Brosens, Koenraad and Verbert, Katrien},
	year = {2020},
	note = {Publisher: IEEE},
	pages = {1084--1094},
}

@article{schich_network_2014,
	title = {A network framework of cultural history},
	volume = {345},
	url = {https://www.science.org/doi/10.1126/science.1240064},
	doi = {10.1126/science.1240064},
	number = {6196},
	urldate = {2024-01-23},
	journal = {Science},
	author = {Schich, Maximilian and Song, Chaoming and Ahn, Yong-Yeol and Mirsky, Alexander and Martino, Mauro and Barabási, Albert-László and Helbing, Dirk},
	month = aug,
	year = {2014},
	note = {Publisher: American Association for the Advancement of Science},
	pages = {558--562},
}

@article{boyd_davis_can_2021,
	title = {Can {I} believe what {I} see? {Data} visualization and trust in the humanities},
	volume = {46},
	issn = {0308-0188},
	shorttitle = {Can {I} believe what {I} see?},
	url = {https://doi.org/10.1080/03080188.2021.1872874},
	doi = {10.1080/03080188.2021.1872874},
	number = {4},
	urldate = {2024-01-23},
	journal = {Interdisciplinary Science Reviews},
	author = {Boyd Davis, Stephen and Vane, Olivia and Kräutli, Florian},
	month = oct,
	year = {2021},
	note = {Publisher: Taylor \& Francis
\_eprint: https://doi.org/10.1080/03080188.2021.1872874},
	pages = {522--546},
}

@article{panagiotidou_communicating_2023,
	title = {Communicating {Uncertainty} in {Digital} {Humanities} {Visualization} {Research}},
	volume = {29},
	issn = {1941-0506},
	doi = {10.1109/TVCG.2022.3209436},
	number = {1},
	journal = {IEEE Transactions on Visualization and Computer Graphics},
	author = {Panagiotidou, Georgia and Lamqaddam, Houda and Poblome, Jeroen and Brosens, Koenraad and Verbert, Katrien and Vande Moere, Andrew},
	month = jan,
	year = {2023},
	pages = {635--645},
}

@inproceedings{janicke_valuable_2016,
	title = {Valuable {Research} for {Visualization} and {Digital} {Humanities}: {A} {Balancing} {Act}},
	booktitle = {Proc. 1st {Workshop} on {Visualization} for the {Digital} {Humanities} ({VIS4DH})},
	author = {Jänicke, Stefan},
	year = {2016},
}

@article{hinrichs_defense_2019,
  title = {In Defense of Sandcastles: {{Research}} Thinking through Visualization in Digital Humanities},
  shorttitle = {In Defense of Sandcastles},
  author = {Hinrichs, Uta and Forlini, Stefania and Moynihan, Bridget},
  year = {2019},
  month = dec,
  journal = {Digital Scholarship in the Humanities},
  volume = {34},
  number = {Supplement\_1},
  pages = {i80-i99},
  issn = {2055-7671},
  doi = {10.1093/llc/fqy051},
  urldate = {2025-09-11}
}

@article{bradley_visualization_2018,
	title = {Visualization and the {Digital} {Humanities}: {Moving} {Toward} {Stronger} {Collaborations}},
	volume = {38},
	shorttitle = {Visualization and the {Digital} {Humanities}},
	doi = {10.1109/MCG.2018.2878900},
	number = {6},
	journal = {IEEE Computer Graphics and Applications},
	author = {Bradley, A. J. and El-Assady, M. and Coles, K. and Alexander, E. and Chen, M. and Collins, C. and Jänicke, S. and Wrisley, D. J.},
	month = nov,
	year = {2018},
	pages = {26--38},
}

@inproceedings{thudt_bohemian_2012,
	address = {New York, NY, USA},
	series = {{CHI} '12},
	title = {The {Bohemian} {Bookshelf}: {Supporting} {Serendipitous} {Book} {Discoveries} {Through} {Information} {Visualization}},
	isbn = {978-1-4503-1015-4},
	shorttitle = {The {Bohemian} {Bookshelf}},
	url = {http://doi.acm.org/10.1145/2207676.2208607},
	doi = {10.1145/2207676.2208607},
	urldate = {2018-09-18},
	booktitle = {Proceedings of the {SIGCHI} {Conference} on {Human} {Factors} in {Computing} {Systems}},
	publisher = {ACM},
	author = {Thudt, Alice and Hinrichs, Uta and Carpendale, Sheelagh},
	year = {2012},
	note = {Citation Key Alias: thudt\_bohemian\_2012-1},
	pages = {1461--1470},
}

@article{whitelaw_generous_2015,
	title = {Generous {Interfaces} for {Digital} {Cultural} {Collections}},
	volume = {009},
	issn = {1938-4122},
	number = {1},
	journal = {Digital Humanities Quarterly},
	author = {Whitelaw, Mitchell},
	month = may,
	year = {2015},
}

@article{mccurdy_poemage_2016,
	title = {Poemage: {Visualizing} the {Sonic} {Topology} of a {Poem}},
	volume = {22},
	issn = {1077-2626},
	shorttitle = {Poemage},
	doi = {10.1109/TVCG.2015.2467811},
	number = {1},
	journal = {IEEE Transactions on Visualization and Computer Graphics},
	author = {McCurdy, N. and Lein, J. and Coles, K. and Meyer, M.},
	month = jan,
	year = {2016},
	note = {Citation Key Alias: mccurdy\_poemage\_2016},
	pages = {439--448},
}

@article{janicke_visualizing_2017,
	title = {Visualizing {Mouvance}: {Toward} a visual analysis of variant medieval text traditions},
	volume = {32},
	issn = {2055-7671},
	shorttitle = {Visualizing {Mouvance}},
	url = {http://www.scopus.com/inward/record.url?scp=85042609030&partnerID=8YFLogxK},
	doi = {10.1093/llc/fqx033},
	urldate = {2023-12-05},
	journal = {Digital Scholarship in the Humanities},
	author = {Jänicke, Stefan and Wrisley, David Joseph},
	month = dec,
	year = {2017},
	pages = {ii106--ii123},
}

@article{koch_varifocalreader_2014,
	title = {{VarifocalReader} — {In}-{Depth} {Visual} {Analysis} of {Large} {Text} {Documents}},
	volume = {20},
	issn = {1077-2626},
	doi = {10.1109/TVCG.2014.2346677},
	number = {12},
	journal = {IEEE Transactions on Visualization and Computer Graphics},
	author = {Koch, S. and John, M. and Wörner, M. and Müller, A. and Ertl, T.},
	month = dec,
	year = {2014},
	pages = {1723--1732},
}

@article{alharbi_transvis_2022,
	title = {{TransVis}: {Integrated} {Distant} and {Close} {Reading} of {Othello} {Translations}},
	volume = {28},
	issn = {1941-0506},
	shorttitle = {{TransVis}},
	url = {https://ieeexplore.ieee.org/abstract/document/9152170?casa_token=tmzzgc2durEAAAAA:Mn5mfYMsIlo-_zeOgZfNYRvhXO6U1k1QQbrWGaskFa1dF9139YXrc1XY_b7NR9XlQAU1kPHxHA},
	doi = {10.1109/TVCG.2020.3012778},
	number = {2},
	urldate = {2023-12-05},
	journal = {IEEE Transactions on Visualization and Computer Graphics},
	author = {Alharbi, Mohammad and Laramee, Robert S and Cheesman, Tom},
	month = feb,
	year = {2022},
	note = {Conference Name: IEEE Transactions on Visualization and Computer Graphics},
	pages = {1397--1414},
}

@article{meinecke_explaining_2022,
	title = {Explaining {Semi}-{Supervised} {Text} {Alignment} {Through} {Visualization}},
	volume = {28},
	issn = {1941-0506},
	url = {https://ieeexplore.ieee.org/document/9516959},
	doi = {10.1109/TVCG.2021.3105899},
	number = {12},
	urldate = {2023-12-05},
	journal = {IEEE Transactions on Visualization and Computer Graphics},
	author = {Meinecke, Christofer and Wrisley, David Joseph and Jänicke, Stefan},
	month = dec,
	year = {2022},
	
	pages = {4797--4809},
}

@inproceedings{baumann_annoxplorer_2020,
	title = {{AnnoXplorer}: {A} {Scalable}, {Integrated} {Approach} for the {Visual} {Analysis} of {Text} {Annotations}},
	volume = {3},
	copyright = {http://creativecommons.org/licenses/by-nc-nd/4.0/},
	isbn = {978-989-758-402-2},
	shorttitle = {{AnnoXplorer}},
	url = {https://www.research-collection.ethz.ch/handle/20.500.11850/386895},
	doi = {10.5220/0008965400620075},
	language = {en},
	urldate = {2023-12-05},
	booktitle = {Proceedings of the 15th {International} {Joint} {Conference} on {Computer} {Vision}, {Imaging} and {Computer} {Graphics} {Theory} and {Applications} ({VISIGRAPP} 2020)},
	publisher = {SciTePress},
	author = {Baumann, Martin and Minasyan, Harutyun and Koch, Steffen and Kurzhals, Kuno and Ertl, Thomas},
	year = {2020},
	pages = {62--75},
}

@inproceedings{sterman_interacting_2020,
	address = {New York, NY, USA},
	title = {Interacting with {Literary} {Style} through {Computational} {Tools}},
	isbn = {978-1-4503-6708-0},
	url = {https://doi.org/10.1145/3313831.3376730},
	urldate = {2022-06-01},
	booktitle = {Proceedings of the 2020 {CHI} {Conference} on {Human} {Factors} in {Computing} {Systems}},
	publisher = {Association for Computing Machinery},
	author = {Sterman, Sarah and Huang, Evey and Liu, Vivian and Paulos, Eric},
	month = apr,
	year = {2020},
	pages = {1--12},
}

@article{geng_shakervis_2015,
	title = {{ShakerVis}: {Visual} analysis of segment variation of {German} translations of {Shakespeare}’s {Othello}},
	volume = {14},
	issn = {1473-8716},
	shorttitle = {{ShakerVis}},
	url = {https://doi.org/10.1177/1473871613495845},
	doi = {10.1177/1473871613495845},
	language = {en},
	number = {4},
	urldate = {2018-10-26},
	journal = {Information Visualization},
	author = {Geng, Zhao and Cheesman, Tom and Laramee, Robert S. and Flanagan, Kevin and Thiel, Stephan},
	month = oct,
	year = {2015},
	pages = {273--288},
}

@article{janicke_visual_2017,
	title = {Visual {Text} {Analysis} in {Digital} {Humanities}},
	volume = {36},
	copyright = {© 2016 The Authors Computer Graphics Forum © 2016 The Eurographics Association and John Wiley \& Sons Ltd.},
	issn = {1467-8659},
	url = {https://onlinelibrary.wiley.com/doi/abs/10.1111/cgf.12873},
	doi = {10.1111/cgf.12873},
	number = {6},
	urldate = {2018-09-18},
	journal = {Computer Graphics Forum},
	author = {Jänicke, S. and Franzini, G. and Cheema, M. F. and Scheuermann, G.},
	month = sep,
	year = {2017},
	pages = {226--250},
}

@article{isenberg_systematic_2013,
	title = {A {Systematic} {Review} on the {Practice} of {Evaluating} {Visualization}},
	volume = {19},
	issn = {1077-2626},
	doi = {10.1109/TVCG.2013.126},
	number = {12},
	journal = {IEEE Transactions on Visualization and Computer Graphics},
	author = {Isenberg, T. and Isenberg, P. and Chen, J. and Sedlmair, M. and Möller, T.},
	month = dec,
	year = {2013},
	pages = {2818--2827},
}

@inproceedings{janicke_close_2015,
	title = {On {Close} and {Distant} {Reading} in {Digital} {Humanities}: {A} {Survey} and {Future} {Challenges}},
	doi = {10.2312/eurovisstar.20151113},
	booktitle = {Eurographics {Conference} on {Visualization} ({EuroVis}) - {STARs}},
	publisher = {The Eurographics Association},
	author = {Jänicke, Stefan and Franzini, Greta and Cheema, Muhammad Faisal and Scheuermann, Gerik},
	editor = {Borgo, R. and Ganovelli, F. and Viola, I.},
	year = {2015},
}

@Article{drucker_et_al:DagRep.13.9.137,
  author =	{Drucker, Johanna and El-Assady, Mennatallah and Hinrichs, Uta and Windhager, Florian and Akbaba, Derya},
  title =	{{Visualization and the Humanities: Towards a Shared Research Agenda (Dagstuhl Seminar 23381)}},
  pages =	{137--165},
  journal =	{Dagstuhl Reports},
  ISSN =	{2192-5283},
  year =	{2024},
  volume =	{13},
  number =	{9},
  editor =	{Drucker, Johanna and El-Assady, Mennatallah and Hinrichs, Uta and Windhager, Florian and Akbaba, Derya},
  publisher =	{Schloss Dagstuhl -- Leibniz-Zentrum f{\"u}r Informatik},
  address =	{Dagstuhl, Germany},
  URL =		{https://drops.dagstuhl.de/entities/document/10.4230/DagRep.13.9.137},
  URN =		{urn:nbn:de:0030-drops-198246},
  doi =		{10.4230/DagRep.13.9.137}
}

@incollection{borek2021_tadirah,
          author = {Luise Borek and Canan Hastik and Vera Khramova and Klaus Illmayer and Jonathan D. Geiger},
            year = {2021},
           title = {Information Organization and Access in Digital Humanities: TaDiRAH Revised, Formalized and FAIR},
          series = {Schriften zur Informationswissenschaft},
       publisher = {Werner H{\"u}lsbusch},
       booktitle = {Information between Data and Knowledge},
          volume = {74},
         address = {Gl{\"u}ckstadt},
            note = {Session 5: Knowledge Representation},
           pages = {321--332},
             url = {https://epub.uni-regensburg.de/44951/}
}

@article{bench_katherine_2020,
	title = {Katherine {Dunham}'s {Global} {Method} and the {Embodied} {Politics} of {Dance}'s {Everyday}},
	volume = {61},
	copyright = {http://creativecommons.org/licenses/by-nc-nd/4.0/},
	issn = {0040-5574, 1475-4533},
	url = {https://www.cambridge.org/core/product/identifier/S0040557420000253/type/journal_article},
	doi = {10.1017/S0040557420000253},
	language = {en},
	number = {3},
	urldate = {2024-07-28},
	journal = {Theatre Survey},
	author = {Bench, Harmony and Elswit, Kate},
	month = sep,
	year = {2020},
	pages = {305--330},
}

@article{panagiotidou_communicating_2022,
	title = {Communicating qualitative uncertainty in data visualization {Two} cases from within the digital humanities},
	volume = {27},
	issn = {0142-5471},
	doi = {10.1075/idj.22014.pan},
	language = {English},
	number = {1},
	journal = {Information Design Journal},
	author = {Panagiotidou, G. and Moere, A.V.},
	year = {2022},
	pages = {52--63},
}

@inproceedings{john_multicloud_2018,
	address = {Waterloo, CAN},
	series = {{GI} '18},
	title = {{MultiCloud}: {Interactive} {Word} {Cloud} {Visualization} for the {Analysis} of {Multiple} {Texts}},
	isbn = {978-0-9947868-3-8},
	shorttitle = {{MultiCloud}},
	url = {https://doi.org/10.20380/GI2018.06},
	doi = {10.20380/GI2018.06},
	urldate = {2024-07-05},
	booktitle = {Proceedings of the 44th {Graphics} {Interface} {Conference}},
	publisher = {Canadian Human-Computer Communications Society},
	author = {John, Markus and Marbach, Eduard and Lohmann, Steffen and Heimerl, Florian and Ertl, Thomas},
	month = jun,
	year = {2018},
	pages = {34--41},
}

@article{han_hisva_2022,
	title = {{HisVA}: {A} {Visual} {Analytics} {System} for {Studying} {History}},
	volume = {28},
	issn = {1941-0506},
	shorttitle = {{HisVA}},
	url = {https://ieeexplore.ieee.org/abstract/document/9447222?casa_token=tumAmputz6sAAAAA:jDpsnBk2wUiGKd_nt8ZyzwDl6wckYfeg4isj93s_5b5NmwWGd6KkhpMu5SU3E9w1n_FY8nXZ},
	doi = {10.1109/TVCG.2021.3086414},
	number = {12},
	urldate = {2024-03-27},
	journal = {IEEE Transactions on Visualization and Computer Graphics},
	author = {Han, Dongyun and Parsad, Gorakh and Kim, Hwiyeon and Shim, Jaekyom and Kwon, Oh-Sang and Son, Kyung A. and Lee, Jooyoung and Cho, Isaac and Ko, Sungahn},
	month = dec,
	year = {2022},	
	pages = {4344--4359},
}

@article{mercun_presenting_2017,
	title = {Presenting bibliographic families using information visualization: {Evaluation} of {FRBR}-based prototype and hierarchical visualizations},
	volume = {68},
	copyright = {© 2016 ASIS\&T},
	issn = {2330-1643},
	shorttitle = {Presenting bibliographic families using information visualization},
	url = {https://onlinelibrary.wiley.com/doi/abs/10.1002/asi.23659},
	doi = {10.1002/asi.23659},
	language = {en},
	number = {2},
	urldate = {2024-03-26},
	journal = {Journal of the Association for Information Science and Technology},
	author = {Merčun, Tanja and Žumer, Maja and Aalberg, Trond},
	year = {2017},
	
	pages = {392--411},
}

@article{benito-santos_evaluating_2021,
	title = {Evaluating a {Taxonomy} of {Textual} {Uncertainty} for {Collaborative} {Visualisation} in the {Digital} {Humanities}},
	volume = {12},
	copyright = {http://creativecommons.org/licenses/by/3.0/},
	url = {https://www.mdpi.com/2078-2489/12/11/436},
	doi = {10.3390/info12110436},
	language = {en},
	number = {11},
	urldate = {2021-10-21},
	journal = {Information},
	author = {Benito-Santos, Alejandro and Doran, Michelle and Rocha, Aleyda and Wandl-Vogt, Eveline and Edmond, Jennifer and Therón, Roberto},
	month = nov,
	year = {2021},
	note = {Number: 11
Publisher: Multidisciplinary Digital Publishing Institute},
	pages = {436},
}

@article{bardiot_theatre_2020,
	title = {Theatre analytics: developing software for theatre research},
	volume = {14},
	issn = {1938-4122},
	shorttitle = {Theatre analytics},
	language = {English},
	number = {3},
	journal = {Digital Humanities Quarterly},
	author = {Bardiot, C.},
	year = {2020},
}

@article{kimura_visualization_2013,
	title = {Visualization of relationships among historical persons from {Japanese} historical documents},
	volume = {28},
	issn = {0268-1145, 1477-4615},
	url = {https://academic.oup.com/dsh/article-lookup/doi/10.1093/llc/fqs045},
	doi = {10.1093/llc/fqs045},
	language = {en},
	number = {2},
	urldate = {2024-03-14},
	journal = {Literary and Linguistic Computing},
	author = {Kimura, F. and Osaki, T. and Tezuka, T. and Maeda, A.},
	month = jun,
	year = {2013},
	pages = {271--278},
}

@book{drucker_visualization_2020,
	title = {Visualization and interpretation: {Humanistic} approaches to display},
	shorttitle = {Visualization and interpretation},
	url = {https://books.google.com/books?hl=en&lr=&id=Ts_tDwAAQBAJ&oi=fnd&pg=PP8&dq=johanna+drucker+visualization&ots=cTmKF3UftQ&sig=QqXHx393_S6XUlkUFvLk4S6wF9Y},
	urldate = {2024-02-15},
	publisher = {MIT Press},
	author = {Drucker, Johanna},
	year = {2020},
}

@article{drucker_humanities_2011,
	title = {Humanities {Approaches} to {Graphical} {Display}},
	volume = {5},
	issn = {1938-4122},
	url = {https://pdfs.semanticscholar.org/e0fe/227ff7a3822f5c0bd41cc566f1a472cc22f2.pdf},
	number = {1},
	journal = {Digital Humanities Quarterly},
	author = {Drucker, Johanna},
	year = {2011},
	note = {Citation Key Alias: drucker\_humanities\_2011, drucker\_humanities\_2011-3},
	pages = {1--21},
}

@article{janicke_interactive_2016,
	title = {Interactive {Visual} {Profiling} of {Musicians}},
	volume = {22},
	issn = {1077-2626},
	doi = {10.1109/TVCG.2015.2467620},
	number = {1},
	journal = {IEEE Transactions on Visualization and Computer Graphics},
	author = {Jänicke, S. and Focht, J. and Scheuermann, G.},
	month = jan,
	year = {2016},
	pages = {200--209},
}

@article{hinrichs_speculative_2016,
	title = {Speculative {Practices}: {Utilizing} {InfoVis} to {Explore} {Untapped} {Literary} {Collections}},
	volume = {22},
	issn = {1077-2626},
	shorttitle = {Speculative {Practices}},
	doi = {10.1109/TVCG.2015.2467452},
	number = {1},
	journal = {IEEE Transactions on Visualization and Computer Graphics},
	author = {Hinrichs, U. and Forlini, S. and Moynihan, B.},
	month = jan,
	year = {2016},
	note = {Citation Key Alias: hinrichs\_speculative\_2016},
	pages = {429--438},
}

@article{zhang_cohortva_2022,
	title = {{CohortVA}: {A} {Visual} {Analytic} {System} for {Interactive} {Exploration} of {Cohorts} {Based} on {Historical} {Data}},
	issn = {1077-2626, 1941-0506, 2160-9306},
	shorttitle = {{CohortVA}},
	url = {https://ieeexplore.ieee.org/document/9912359/},
	doi = {10.1109/TVCG.2022.3209483},
	urldate = {2023-11-14},
	journal = {IEEE Transactions on Visualization and Computer Graphics},
	author = {Zhang, Wei and Wong, Jason K. and Wang, Xumeng and Gong, Youcheng and Zhu, Rongchen and Liu, Kai and Yan, Zihan and Tan, Siwei and Qu, Huamin and Chen, Siming and Chen, Wei},
	year = {2022},
	pages = {1--11},
}

@article{miller_corpusvis_2022,
	title = {{CorpusVis}: {Visual} {Analysis} of {Digital} {Sheet} {Music} {Collections}},
	volume = {41},
	copyright = {© 2022 The Author(s). Computer Graphics Forum published by Eurographics - The European Association for Computer Graphics and John Wiley \& Sons Ltd.},
	issn = {1467-8659},
	shorttitle = {{CorpusVis}},
	url = {https://onlinelibrary.wiley.com/doi/abs/10.1111/cgf.14540},
	doi = {10.1111/cgf.14540},
	language = {en},
	number = {3},
	urldate = {2023-11-21},
	journal = {Computer Graphics Forum},
	author = {Miller, Matthias and Rauscher, Julius and Keim, Daniel A. and El-Assady, Mennatallah},
	year = {2022},
	pages = {283--294},
}

@article{meinecke_towards_2022,
	title = {Towards {Enhancing} {Virtual} {Museums} by {Contextualizing} {Art} through {Interactive} {Visualizations}},
	volume = {15},
	issn = {1556-4673},
	url = {https://dl.acm.org/doi/10.1145/3527619},
	doi = {10.1145/3527619},
	number = {4},
	urldate = {2023-12-05},
	journal = {Journal on Computing and Cultural Heritage},
	author = {Meinecke, Christofer and Hall, Chris and Jänicke, Stefan},
	month = dec,
	year = {2022},
	pages = {62:1--62:26},
}

@article{liu_storyflow_2013,
	title = {{StoryFlow}: {Tracking} the {Evolution} of {Stories}},
	volume = {19},
	issn = {1077-2626},
	shorttitle = {{StoryFlow}},
	doi = {10.1109/TVCG.2013.196},
	number = {12},
	journal = {IEEE Transactions on Visualization and Computer Graphics},
	author = {Liu, S. and Wu, Y. and Wei, E. and Liu, M. and Liu, Y.},
	month = dec,
	year = {2013},
	pages = {2436--2445},
}

@article{windhager_visualization_2018,
	title = {Visualization of {Cultural} {Heritage} {Collection} {Data}: {State} of the {Art} and {Future} {Challenges}},
	issn = {1077-2626},
	shorttitle = {Visualization of {Cultural} {Heritage} {Collection} {Data}},
	doi = {10.1109/TVCG.2018.2830759},
	journal = {IEEE Transactions on Visualization and Computer Graphics},
	author = {Windhager, F. and Federico, P. and Schreder, G. and Glinka, K. and Dörk, M. and Miksch, S. and Mayr, E.},
	year = {2018},
	pages = {1--1},
}

@article{benito-santos_data-driven_2020,
	title = {A {Data}-{Driven} {Introduction} to {Authors}, {Readings}, and {Techniques} in {Visualization} for the {Digital} {Humanities}},
	volume = {40},
	copyright = {All rights reserved},
	issn = {1558-1756},
	doi = {10.1109/MCG.2020.2973945},
	number = {3},
	journal = {IEEE Computer Graphics and Applications},
	author = {Benito-Santos, A. and Therón Sánchez, Roberto},
	month = may,
	year = {2020},
	pages = {45--57},
}

@article{summative,
  title={The validity, generalizability and feasibility of summative evaluation methods in visual analytics},
  author={Khayat, Mosab and Karimzadeh, Morteza and Ebert, David S and Ghafoor, Arif},
  journal={IEEE Transactions on Visualization and Computer Graphics},
  volume={26},
  number={1},
  pages={353--363},
  year={2019},
  publisher={IEEE}
}

@incollection{Carpendale2008,
    author={Carpendale, Sheelagh},
    editor={Kerren, Andreas and Stasko, John T. and Fekete, Jean-Daniel and North, Chris},
    title={Evaluating Information Visualizations},
    booktitle={Information Visualization: Human-Centered Issues and Perspectives},
    year={2008},
    publisher={Springer Berlin Heidelberg},
    address={Berlin, Heidelberg},
    pages={19--45},
    isbn={978-3-540-70956-5},
    doi={10.1007/978-3-540-70956-5_2},
}

@inproceedings{Isenberg2008,
author = {Isenberg, Petra and Zuk, Torre and Collins, Christopher and Carpendale, Sheelagh},
    title = {Grounded evaluation of information visualizations},
    year = {2008},
    isbn = {9781605580166},
    hide_publisher = {Association for Computing Machinery},
    hide_address = {New York, NY, USA},
    doi = {10.1145/1377966.1377974},
    booktitle = {Proc. Workshop on BEyond Time and Errors: Novel EvaLuation Methods for Information Visualization (BELIV)},
    articleno = {6},
    numpages = {8},
    location = {Florence, Italy},
    hide_series = {BELIV '08}
}

@inproceedings{Plaisant2004,
    author = {Plaisant, Catherine},
    title = {The challenge of information visualization evaluation},
    year = {2004},
    isbn = {1581138679},
    hide_publisher = {Association for Computing Machinery},
    hide_address = {New York, NY, USA},
    doi = {10.1145/989863.989880},
    booktitle = {Proceedings of the Working Conference on Advanced Visual Interfaces (AVI)},
    pages = {109–116},
    numpages = {8},
    location = {Gallipoli, Italy},
    hide_series = {AVI '04}
}

@ARTICLE{North2006,
  author={North, C.},
  journal={IEEE Computer Graphics \& Applications}, 
  title={Toward measuring visualization insight}, 
  year={2006},
  volume={26},
  number={3},
  pages={6-9},
  doi={10.1109/MCG.2006.70}}

@inproceedings{dal2002evaluating,
    title={Evaluating usability of information visualization techniques},
    author = {Dal, Carla and Freitas, Carla and Luzzardi, Paulo and Cava, Ricardo and Winckler, Marco and Pimenta, Marcelo and Nedel, Luciana},
    booktitle={Brazilian Symposium on Human Factors in Computing Systems},
    year={2002}
}

@book{Banissi2014,
    author = {Banissi, E. (Ebad)},
    address = {Newcastle upon Tyne},
    booktitle = {Information visualisation : techniques, usability and evaluation},
    edition = {1st ed.},
    language = {eng},
    publisher = {Cambridge Scholars Publishing},
    title = {Information visualisation : techniques, usability and evaluation },
    year = {2014},
    isbn = {1-4438-7017-X},
}

@inproceedings{saket2016beyond,
    author = {Saket, Bahador and Endert, Alex and Stasko, John},
    title = {Beyond Usability and Performance: A Review of User Experience-Focused Evaluations in Visualization},
    year = {2016},
    isbn = {9781450348188},
    hide_publisher = {Association for Computing Machinery},
    hide_address = {New York, NY, USA},
    doi = {10.1145/2993901.2993903},
    booktitle = {Proceedings of the Sixth Workshop on Beyond Time and Errors on Novel Evaluation Methods for Visualization (BELIV)},
    pages = {133–142},
    numpages = {10},
    hide_location = {Baltimore, MD, USA},
    hide_series = {BELIV '16}
}

@inproceedings{stasko2014value,
    author = {Stasko, John},
    title = {Value-driven evaluation of visualizations},
    year = {2014},
    isbn = {9781450332095},
    publisher = {Association for Computing Machinery},
    address = {New York, NY, USA},
    doi = {10.1145/2669557.2669579},
    booktitle = {Proceedings of the Fifth Workshop on Beyond Time and Errors: Novel Evaluation Methods for Visualization},
    pages = {46–53},
    numpages = {8},
    location = {Paris, France},
    series = {BELIV '14}
}

@ARTICLE{saraiya2006,
  author={Saraiya, P. and North, C. and Vy Lam and Duca, K.A.},
  journal={IEEE Trans. Visualization \& Computer Graphics}, 
  title={An Insight-Based Longitudinal Study of Visual Analytics}, 
  year={2006},
  volume={12},
  number={6},
  pages={1511-1522},
  doi={10.1109/TVCG.2006.85}}

@ARTICLE{Smuc2009,
  author={Smuc, Michael and Mayr, Eva and Lammarsch, Tim and Aigner, Wolfgang and Miksch, Silvia and Gärtner, Johannes},
  journal={IEEE Computer Graphics \& Applications}, 
  title={To Score or Not to Score? {Tripling} Insights for Participatory Design}, 
  year={2009},
  volume={29},
  number={3},
  pages={29-38},
  doi={10.1109/MCG.2009.53}}

@inproceedings{crisan2018how,
  author={Crisan, Anamaria and Elliott, Madison},
  booktitle={2018 IEEE Evaluation and Beyond - Methodological Approaches for Visualization (BELIV)}, 
  title={How to Evaluate an Evaluation Study? Comparing and Contrasting Practices in Vis with Those of Other Disciplines : Position Paper}, 
  year={2018},
  volume={},
  number={},
  pages={28-36},
  doi={10.1109/BELIV.2018.8634420}
}

@inproceedings{Preim2018ACA,
  title={A Critical Analysis of the Evaluation Practice in Medical Visualization},
  author={B. Preim and Timo Ropinski and Petra Isenberg},
  booktitle={VCBM@MICCAI},
  year={2018},
  doi={10.2312/vcbm.20181228}
}

@ARTICLE{Windhager:TVCG:2019,
  author={Windhager, Florian and Federico, Paolo and Schreder, Günther and Glinka, Katrin and Dörk, Marian and Miksch, Silvia and Mayr, Eva},
  journal={IEEE Transactions on Visualization and Computer Graphics}, 
  title={Visualization of Cultural Heritage Collection Data: State of the Art and Future Challenges}, 
  year={2019},
  volume={25},
  number={6},
  pages={2311-2330},
  doi={10.1109/TVCG.2018.2830759}}

@ARTICLE{Sedlmair:TVCG:2012,
  author={Sedlmair, Michael and Meyer, Miriah and Munzner, Tamara},
  journal={IEEE Transactions on Visualization and Computer Graphics}, 
  title={Design Study Methodology: Reflections from the Trenches and the Stacks}, 
  year={2012},
  volume={18},
  number={12},
  pages={2431-2440},
  doi={10.1109/TVCG.2012.213}}

@article{Hinrichs:DSH:2018,
    author = {Hinrichs, Uta and Forlini, Stefania and Moynihan, Bridget},
    title = {In defense of sandcastles: Research thinking through visualization in digital humanities},
    journal = {Digital Scholarship in the Humanities},
    volume = {34},
    number = {Supplement 1},
    pages = {80-99},
    year = {2018},
    month = {10},
    issn = {2055-7671},
    doi = {10.1093/llc/fqy051}
}

@ARTICLE{lin_hunches,
  author={Lin, Haihan and Akbaba, Derya and Meyer, Miriah and Lex, Alexander},
  journal={IEEE Transactions on Visualization and Computer Graphics}, 
  title={Data Hunches: Incorporating Personal Knowledge into Visualizations}, 
  year={2023},
  volume={29},
  number={1},
  pages={504-514},
  doi={10.1109/TVCG.2022.3209451}}

@ARTICLE{perovich,
  author={Perovich, Laura J. and Wylie, Sara Ann and Bongiovanni, Roseann},
  journal={IEEE Transactions on Visualization and Computer Graphics}, 
  title={Chemicals in the Creek: designing a situated data physicalization of open government data with the community}, 
  year={2021},
  volume={27},
  number={2},
  pages={913-923},
  doi={10.1109/TVCG.2020.3030472}}

@inproceedings{schmidt,
  title={A public exploratory data analysis of gender bias in teaching evaluations},
  author={Schmidt, Benajmin},
  booktitle={vis4dh Proceedings},
  year={2016}
}

@article{bentkowska2013bought,
  title={“I bought a piece of Roman furniture on the Internet. It's quite good but low on polygons.”—Digital Visualization of Cultural Heritage and its Scholarly Value in Art History},
  author={Bentkowska-Kafel, Anna},
  journal={Visual Resources},
  volume={29},
  number={1-2},
  pages={38--46},
  year={2013},
  publisher={Taylor \& Francis}
}

@misc{correll2019counting,
    Author = {Correll, Michael},
    Howpublished = {Blogpost},
    Month = {09},
    Title = {Counting, Collaborating, and Coexisting: Visualization and the Digital Humanities},
    Year = {2019},
    URL = {https://mcorrell.medium.com/counting-collaborating-and-coexisting-visualization-and-the-digital-humanities-1bf157400d8}
}

@article{cheesman2017multi,
  title={Multi-retranslation corpora: Visibility, variation, value, and virtue},
  author={Cheesman, Tom and Flanagan, Kevin and Thiel, Stephan and Rybicki, Jan and Laramee, Robert S and Hope, Jonathan and Roos, Avraham},
  journal={Digital Scholarship in the Humanities},
  volume={32},
  number={4},
  pages={739--760},
  year={2017},
  publisher={Oxford University Press}
}

@article{schwan_disclosure,
  title={Disclosure as a critical-feminist design practice for Web-based data stories},
  author={Schwan, Hannah and Arndt, Jonas and D{\"o}rk, Marian},
  journal={First Monday},
  year={2022}
}

@article{bornhofen,
  title={Exploring dynamic multilayer graphs for digital humanities},
  author={Bornhofen, Stefan and D{\"u}ring, Marten},
  journal={Applied Network Science},
  volume={5},
  number={1},
  pages={54},
  year={2020},
  publisher={Springer}
}

@article{chen_hierarchical,
  title={A hierarchical topic analysis tool to facilitate digital humanities research},
  author={Chen, Chih-Ming and Ho, Szu-Yu and Chang, Chung},
  journal={Aslib Journal of Information Management},
  volume={75},
  number={1},
  pages={1--19},
  year={2023},
  publisher={Emerald Publishing Limited}
}

@inproceedings{mayr2022multiple,
  title={The multiple faces of cultural heritage: Towards an integrated visualization platform for tangible and intangible cultural assets},
  author={Mayr, Eva and Windhager, Florian and Liem, Johannes and Beck, Samuel and Koch, Steffen and Kusnick, Jakob and J{\"a}nicke, Stefan},
  booktitle={2022 IEEE 7th Workshop on Visualization for the Digital Humanities (VIS4DH)},
  pages={13--18},
  year={2022},
  organization={IEEE}
}

@book{munzner2014visualization,
  title={Visualization analysis and design},
  author={Munzner, Tamara},
  year={2014},
  publisher={CRC press}
}

@article{during2020what_were,
  title={What Were the Humanities, Anyway? This dangerous moment demands that we give an elusive concept its history},
  author={During, Simon},
  journal={The Chronicle of Higher Education, August},
  volume={31},
  year={2020}
}

@article{windhager2024complexity,
  title={Complexity as Design Material},
  author={Windhager, Florian and Abduhl-Rahman, Alfie and Bludau, Mark-Jan and Hengesbach, Nicole and Lamqaddam, Houda and Meirelles, Isabel and Speckmann, Bettina and Correll, Michael},
  journal={arXiv preprint arXiv:2409.07465},
  year={2024}
}

@article{windhager2024DH_and_DC,
  title={Digital Humanities and Distributed Cognition: From a Lack of Theory to its Visual Augmentation},
  author={Windhager, Florian and Mayr, Eva},
  journal={Journal of Cultural Analytics},
  volume={7},
  number={4},
  year={2024}
}

@article{kauer2024discursive,
  title={Discursive Patinas: Anchoring Discussions in Data Visualizations},
  author={Kauer, Tobias and Akbaba, Derya and D{\"o}rk, Marian and Bach, Benjamin},
  journal={IEEE transactions on visualization and computer graphics},
  year={2024},
  publisher={IEEE}
}

@article{jackson2024worldwide,
  title={Worldwide divergence of values},
  author={Jackson, Joshua Conrad and Medvedev, Danila},
  journal={Nature Communications},
  volume={15},
  number={1},
  pages={2650},
  year={2024},
  publisher={Nature Publishing Group UK London}
}

@book{inglehart2005christian,
  title = {Modernization, {{Cultural Change}}, and {{Democracy}}: {{The Human Development Sequence}}},
  shorttitle = {Modernization, {{Cultural Change}}, and {{Democracy}}},
  author = {Inglehart, Ronald and Welzel, Christian},
  year = {2005},
  publisher = {Cambridge University Press},
  address = {Cambridge},
  doi = {10.1017/CBO9780511790881},
  urldate = {2025-09-11},
  isbn = {978-0-521-84695-0}
}

@Misc{icomos1982florence-charta,
  author = {{International Council on Monuments and Sites}},
  title  = {Historic Gardens, {The} {Florence} Charter, {Florence}, {Italy}},
  year   = {1982},
}

@INPROCEEDINGS{Lin2024,
  author={Lin, Feng and Wang, Arran Zeyu and Rahman, Md Dilshadur and Szafir, Danielle Albers and Quadri, Ghulam Jilani},
  booktitle={2024 IEEE Evaluation and Beyond - Methodological Approaches for Visualization (BELIV)}, 
  title={Striking the Right Balance: Systematic Assessment of Evaluation Method Distribution Across Contribution Types}, 
  year={2024},
  volume={},
  number={},
  pages={129-135},
  doi={10.1109/BELIV64461.2024.00020}}

@article{xing_review_2025,
  title = {A {{Review}} and {{Analysis}} of {{Evaluation Practices}} in {{VIS Domain Applications}}},
  author = {Xing, Yiwen and Cantareira, Gabriel D. and Borgo, Rita and {Abdul-Rahman}, Alfie},
  year = 2025,
  month = sep,
  journal = {IEEE Transactions on Visualization and Computer Graphics},
  volume = {31},
  number = {9},
  pages = {5580--5592},
  issn = {1941-0506},
  doi = {10.1109/TVCG.2024.3460181},
  urldate = {2025-08-25}
}

@article{meyer2019criteria,
  title={Criteria for rigor in visualization design study},
  author={Meyer, Miriah and Dykes, Jason},
  journal={IEEE transactions on visualization and computer graphics},
  volume={26},
  number={1},
  pages={87--97},
  year={2019},
  publisher={IEEE}
}

@article{berret_iceberg_2025,
  title = {Iceberg {{Sensemaking}}: {{A Process Model}} for {{Critical Data Analysis}}},
  shorttitle = {Iceberg {{Sensemaking}}},
  author = {Berret, Charles and Munzner, Tamara},
  year = 2025,
  month = sep,
  journal = {IEEE Transactions on Visualization and Computer Graphics},
  volume = {31},
  number = {9},
  pages = {6067--6084},
  issn = {1941-0506},
  doi = {10.1109/TVCG.2024.3486613},
  urldate = {2025-11-14}
}

@article{windhager2025exploration,
  title={From Exploration to Critique: Catalyzing Critical Inquiry with Cultural Collection Visualizations},
  author={Windhager, Florian and Mayr, Eva and Glinka, Katrin},
  journal={IEEE Computer Graphics and Applications},
  year={2025},
  publisher={IEEE}
}

@book{hall2022critical,
  title={Critical Visualization: Rethinking the Representation of Data},
  author={Hall, Peter A and D{\'a}vila, Patricio},
  year={2022},
  publisher={Bloomsbury Publishing}
}

@article{dagstuhl_2024_vis4dh,
	title = {Visualization and the {Humanities}: {Towards} a {Shared} {Research} {Agenda} (23381)},
	copyright = {https://creativecommons.org/licenses/by/4.0/legalcode},
	shorttitle = {Visualization and the {Humanities}},
	urldate = {2024-06-27},
	journal = {Dagstuhl Reports},
        volume = {13},
        issue ={9},
	author = {Drucker, Johanna and El-Assady, Mennatallah and Hinrichs, Uta and Windhager, Florian and Akbaba, Derya},
	year = {2024},
}

@article{warwick2015building,
  title={Building theories or theories of building? A tension at the heart of digital humanities},
  author={Warwick, Claire},
  journal={A new companion to digital humanities},
  pages={538--552},
  year={2015},
  publisher={Wiley Online Library}
}

@article{heinicker2023more,
  title={More than Distant Viewing: Qualitative Views on Machine Learning as an Automated Analysis Method in Networked Climate Image Communication.},
  author={Heinicker, Paul and Kienbaum, Janna and Schneider, Birgit},
  journal={DHQ: Digital Humanities Quarterly},
  volume={17},
  number={1},
  year={2023}
}

@article{moretti2016alcide,
  title={ALCIDE: Extracting and visualising content from large document collections to support humanities studies},
  author={Moretti, Giovanni and Sprugnoli, Rachele and Menini, Stefano and Tonelli, Sara},
  journal={Knowledge-Based Systems},
  volume={111},
  pages={100--112},
  year={2016},
  publisher={Elsevier}
}

@inproceedings{roessler2019textiles,
  title={TexTiles: Exploring Patterns in Historical Discourse},
  author={Roessler, Robert and Kelly, Caiseen and Behrisch, Michael and Beyer, Johanna},
  booktitle={4th Workshop on Visualization for the Digital Humanities, IEEE VIS},
  year={2019}
}

@article{rybenska2024review,
  title={Review of Modern Approaches to 3D Digitization of Tangible Cultural Heritage},
  author={Rybenská, Klára and Borůvková, Barbora},
  journal={Journal of Digital Art and Humanities},
  volume={5},
  number={1},
  pages={20--30},
  year={2024},
  publisher={Institute of Cited Scientists}
}

@article{koller2010research,
  title={Research challenges for digital archives of 3D cultural heritage models},
  author={Koller, David and Frischer, Bernard and Humphreys, Greg},
  journal={Journal on Computing and Cultural Heritage (JOCCH)},
  volume={2},
  number={3},
  pages={1--17},
  year={2010},
  publisher={ACM New York, NY, USA}
}

@inproceedings{hengesbach2022undoing,
  title={Undoing seamlessness: Exploring seams for critical visualization},
  author={Hengesbach, Nicole},
  booktitle={CHI conference on human factors in computing systems extended abstracts},
  pages={1--7},
  year={2022}
}

@book{everitt_cluster_2011,
  title={Cluster Analysis},
  author={Everitt, Brian S. and Landau, Sabine and Leese, Morven and Stahl, Daniel},
  year={2011},
  publisher={John Wiley \& Sons},
  edition={5th}
}

@article{ward_hierarchical_1963,
  title={Hierarchical Grouping to Optimize an Objective Function},
  author={Ward, Jr., Joe H.},
  journal={Journal of the American Statistical Association},
  volume={58},
  number={301},
  pages={236--244},
  year={1963},
  publisher={Taylor \& Francis}
}

@book{hastie_elements_2009,
  title={The Elements of Statistical Learning: Data Mining, Inference, and Prediction},
  author={Hastie, Trevor and Tibshirani, Robert and Friedman, Jerome},
  year={2009},
  publisher={Springer Science \& Business Media},
  edition={2nd}
}

@article{vancisin2023provenance,
  title={Provenance visualization: Tracing people, processes, and practices through a data-driven approach to provenance},
  author={Vancisin, Tomas and Clarke, Loraine and Orr, Mary and Hinrichs, Uta},
  journal={Digital Scholarship in the Humanities},
  volume={38},
  number={3},
  pages={1322--1339},
  year={2023},
  publisher={Oxford University Press}
}

@inproceedings{wrisley_pre-visualization_2018,
    title = {Pre-visualization},
    language = {English},
    booktitle = {Proc. 3rd {Workshop} on {Visualization} for the {Digital} {Humanities} ({VIS4DH})},
    author = {Wrisley, David Joseph},
    year = {2018},
}

@inproceedings{hinrichs2017risk,
  title={Risk the drift! Stretching disciplinary boundaries through critical collaborations between the humanities and visualization},
  author={Hinrichs, Uta and El-Assady, Mennatallah and Bradely, Adam James and Forlini, Stefania and Collins, Christopher},
booktitle={Proc. 2nd Workshop of Visualization for the Digital Humanities (VIS4DH)},
  year={2017}
}

@article{hall2019design,
  title={Design by immersion: A transdisciplinary approach to problem-driven visualizations},
  author={Hall, Kyle Wm and Bradley, Adam J and Hinrichs, Uta and Huron, Samuel and Wood, Jo and Collins, Christopher and Carpendale, Sheelagh},
  journal={IEEE transactions on visualization and computer graphics},
  volume={26},
  number={1},
  pages={109--118},
  year={2019},
  publisher={IEEE}
}

@inproceedings{krautli2013known,
  title={Known unknowns: Representing uncertainty in historical time},
  author={Kr{\"a}utli, Florian and Davis, Stephen Boyd},
  booktitle={Electronic Visualisation and the Arts (EVA 2013)},
  year={2013},
  organization={BCS Learning \& Development}
}

@inproceedings{bludau2025fluidly,
  title={Fluidly Revealing Information: A Survey of Un/foldable Data Visualizations},
  author={Bludau, M-J and D{\"o}rk, Marian and Bruckner, Stefan and Tominski, Christian},
  booktitle={Computer Graphics Forum},
  pages={e70152},
  year={2025},
  organization={Wiley Online Library}
}

@article{hullman2018pursuit,
  title={In pursuit of error: A survey of uncertainty visualization evaluation},
  author={Hullman, Jessica and Qiao, Xiaoli and Correll, Michael and Kale, Alex and Kay, Matthew},
  journal={IEEE transactions on visualization and computer graphics},
  volume={25},
  number={1},
  pages={903--913},
  year={2018},
  publisher={IEEE}
}

@inproceedings{windhager2019uncertainty_of_what,
  title={Uncertainty of what and for whom-and does anyone care? Propositions for cultural collection visualization},
  author={Windhager, Florian and Salisu, Saminu and Schreder, G{\"u}nther and Mayr, Eva},
  booktitle={4th IEEE Workshop on Visualization for the Digital Humanities (VIS4DH)},
  pages={1--5},
  year={2019}
}

@article{massari2025representing,
  title={Representing provenance and track changes of cultural heritage metadata in RDF: a survey of existing approaches},
  author={Massari, Arcangelo and Peroni, Silvio and Tomasi, Francesca and Heibi, Ivan},
  journal={Digital Scholarship in the Humanities},
  pages={fqaf076},
  year={2025},
  publisher={Oxford University Press}
}

@incollection{bianco2012digital,
  title={This digital humanities which is not one},
  author={Bianco, Jamie},
  booktitle={Debates in the digital humanities},
  pages={96--112},
  year={2012},
  publisher={University of Minnesota Press}
}

@book{gadamer_truth_2004,
   series = {Continuum impacts},
	title = {Truth and method},
	isbn = {978-0-8264-7697-5},
	language = {eng},
	publisher = {Continuum},
	author = {Gadamer, Hans-Georg and Weinsheimer, Joel and Marshall, Donald G.},
	year = {2004},
}

@book{umberto1994limits,
  title={The limits of interpretation},
  author={Umberto, ECO},
  year={1994},
  publisher={Indiana University Press}
}

@article{latour2012whole,
  title={‘The whole is always smaller than its parts’--a digital test of G abriel T ardes' monads},
  author={Latour, Bruno and Jensen, Pablo and Venturini, Tommaso and Grauwin, S{\'e}bastian and Boullier, Dominique},
  journal={The British journal of sociology},
  volume={63},
  number={4},
  pages={590--615},
  year={2012},
  publisher={Wiley Online Library}
}

@book{nealon2012theory,
  title={The theory toolbox: Critical concepts for the humanities, arts, and social sciences},
  author={Nealon, Jeffrey Thomas and Giroux, Susan Searls},
  year={2012},
  publisher={Rowman \& Littlefield Publishers}
}

@article{bruggemann2020fold,
  title={The fold: Rethinking interactivity in data visualization},
  author={Br{\"u}ggemann, Viktoria and Bludau, Mark-Jan and D{\"o}rk, Marian},
  journal={DHQ},
  volume={14},
  number={3},
  year={2020}
}

@article{nordberg2023lens,
  title={The lens of theory: seeing better or differently?},
  author={Nordberg, Donald},
  journal={Int. J. of Organization Theory \& Behavior},
  volume={26},
  number={1/2},
  pages={152--162},
  year={2023},
  publisher={Emerald Publishing Limited}
}

@incollection{godfrey2009theory,
  title={Theory and reality: An introduction to the philosophy of science},
  author={Godfrey-Smith, Peter},
  booktitle={Theory and reality},
  year={2009},
  publisher={University of Chicago Press}
}

@article{akbaba_entanglements_2025,
  title = {Entanglements for {{Visualization}}: {{Changing Research Outcomes}} through {{Feminist Theory}}},
  shorttitle = {Entanglements for {{Visualization}}},
  author = {Akbaba, Derya and Klein, Lauren and Meyer, Miriah},
  year = 2025,
  month = jan,
  journal = {IEEE Transactions on Visualization and Computer Graphics},
  volume = {31},
  number = {1},
  pages = {1279--1289},
  issn = {1941-0506},
  doi = {10.1109/TVCG.2024.3456171},
  urldate = {2026-01-23}
}

@book{geertz2017interpretation,
  title={The interpretation of cultures},
  author={Geertz, Clifford},
  year={2017},
  publisher={Basic books}
}

@article{liu2012cultural,
  title={Where is cultural criticism in the digital humanities?},
  author={Liu, Alan},
  journal={Debates in the digital humanities},
  volume={2012},
  pages={490--510},
  year={2012},
  publisher={Minneapolis}
}

@book{scott2009ricoeur_hermeneutics_suspicion,
  title={Ricoeur and the Hermeneutics of Suspicion},
  author={Scott-Baumann, Alison},
  year={2009},
  publisher={A\&C Black}
}

@article{panagiotidou2025critical,
  title={Critical Data Visualization—Part I},
  author={Panagiotidou, Georgia and McNutt, Andrew and Akbaba, Derya and Hengesbach, Nicole and Meyer, Miriah},
  journal={IEEE Computer Graphics and Applications},
  volume={45},
  number={3},
  pages={14--16},
  year={2025},
  publisher={IEEE}
}

@article{he2023enthusiastic,
  title={Enthusiastic and grounded, avoidant and cautious: Understanding public receptivity to data and visualizations},
  author={He, Helen Ai and Walny, Jagoda and Thoma, Sonja and Carpendale, Sheelagh and Willett, Wesley},
  journal={IEEE Transactions on Visualization and Computer Graphics},
  volume={30},
  number={1},
  pages={1435--1445},
  year={2023},
  publisher={IEEE}
}

@article{kauer2025towards,
  title={Towards Collective Storytelling: Investigating Audience Annotations in Data Visualizations},
  author={Kauer, Tobias and D{\"o}rk, Marian and Bach, Benjamin},
  journal={IEEE Computer Graphics and Applications},
  year={2025},
  publisher={IEEE}
}

@article{walny2020pixelclipper,
  title={Pixelclipper: Supporting public engagement and conversation about visualizations},
  author={Walny, Jagoda and Storteboom, Sarah and Pusch, Richard and Hwang, Steven Munsu and Knudsen, S{\o}ren and Carpendale, Sheelagh and Willett, Wesley},
  journal={IEEE Computer Graphics and Applications},
  volume={40},
  number={2},
  pages={57--70},
  year={2020},
  publisher={IEEE}
}

@inproceedings{wohlin_guidelines_2014,
  title = {Guidelines for Snowballing in Systematic Literature Studies and a Replication in Software Engineering},
  booktitle = {Proceedings of the 18th International Conference on Evaluation and Assessment in Software Engineering},
  author = {Wohlin, Claes},
  year = {2014},
  pages = {38},
  publisher = {ACM}
}

@inproceedings{dork2011information,
  title={The information flaneur: A fresh look at information seeking},
  author={D{\"o}rk, Marian and Carpendale, Sheelagh and Williamson, Carey},
  booktitle={Proceedings of the SIGCHI conference on human factors in computing systems},
  pages={1215--1224},
  year={2011}
}

@article{taddeo2005solving,
  title={Solving the symbol grounding problem: a critical review of fifteen years of research},
  author={Taddeo, Mariarosaria and Floridi, Luciano},
  journal={Journal of Experimental \& Theoretical Artificial Intelligence},
  volume={17},
  number={4},
  pages={419--445},
  year={2005},
  publisher={Taylor \& Francis}
}

@Manual{polr, title = {polr: Polar OpenLayers}, author = {Ben Raymond}, year = {2025}, note = {R package version 0.0.5}, url = {https://github.com/scar/polr}, }

@article {PRISMA,
    hide_author = {Page, Matthew J and McKenzie, Joanne E and Bossuyt, Patrick M and Boutron, Isabelle and Hoffmann, Tammy C and Mulrow, Cynthia D and Shamseer, Larissa and Tetzlaff, Jennifer M and Akl, Elie A and Brennan, Sue E and Chou, Roger and Glanville, Julie and Grimshaw, Jeremy M and Hr{\'o}bjartsson, Asbj{\o}rn and Lalu, Manoj M and Li, Tianjing and Loder, Elizabeth W and Mayo-Wilson, Evan and McDonald, Steve and McGuinness, Luke A and Stewart, Lesley A and Thomas, James and Tricco, Andrea C and Welch, Vivian A and Whiting, Penny and Moher, David},
    author = {Page, Matthew J and others},
    title = {The {PRISMA} 2020 statement: an updated guideline for reporting systematic reviews},
    volume = {372},
    elocation-id = {n71},
    year = {2021},
    hide_doi = {10.1136/bmj.n71},
    publisher = {BMJ Publishing Group Ltd},
    URL = {https://www.bmj.com/content/372/bmj.n71},
    eprint = {https://www.bmj.com/content/372/bmj.n71.full.pdf},
    journal = {BMJ}
}

@article{kimmel2006culture,
  title={Culture and conflict},
  author={Kimmel, Paul R},
  journal={The handbook of conflict resolution: Theory and practice},
  pages={625--648},
  year={2006},
  publisher={Jossey-Bass San Francisco}
}

\appendix 

\end{document}